\documentclass[10pt, conference, letterpaper]{IEEEtran}
\IEEEoverridecommandlockouts

\usepackage{cite}
\usepackage{amsmath,amssymb,amsfonts}  
\usepackage{algorithmic}
\usepackage{graphicx}
\usepackage{textcomp}
\usepackage{xcolor}
\usepackage[compatibility=false]{caption}
\usepackage{subcaption}
\usepackage{xspace}
\usepackage{gensymb}
\usepackage{url}
\usepackage{hyperref}

\usepackage[T1]{fontenc}
\usepackage{newtxtext,newtxmath}

\newcommand{\sysname}{\textit{ArrayLink}\xspace} 

\def\BibTeX{{\rm B\kern-.05em{\sc i\kern-.025em b}\kern-.08em
    T\kern-.1667em\lower.7ex\hbox{E}\kern-.125emX}}
    
\begin{document}

\title{Satellites are closer than you think: A near field MIMO approach for Ground stations}



\author{
\IEEEauthorblockN{Rohith Reddy Vennam\IEEEauthorrefmark{2},
Luke Wilson\IEEEauthorrefmark{2},
Ish Kumar Jain\IEEEauthorrefmark{4},
Dinesh Bharadia\IEEEauthorrefmark{2}}
\IEEEauthorblockA{\IEEEauthorrefmark{2}University of California San Diego, La Jolla, CA}
\IEEEauthorblockA{\IEEEauthorrefmark{4}Rensselaer Polytechnic Institute, Troy, NY}
\IEEEauthorblockA{Email: rvennam@ucsd.edu, l5wilson@ucsd.edu, jaini@rpi.edu, dineshb@ucsd.edu}
}


\maketitle
\begin{abstract}
The rapid growth of low Earth orbit (LEO) satellite constellations has revolutionized broadband access, Earth observation, and direct-to-device connectivity. However, the expansion of ground station infrastructure has not kept pace, creating a critical bottleneck in satellite-to-ground backhaul capacity. Traditional parabolic dish antennas, though effective for geostationary (GEO) satellites, are ill-suited for dense, fast-moving LEO networks due to mechanical steering delays and their inability to track multiple satellites simultaneously. Phased array antennas offer electronically steerable beams and multi-satellite support, but their integration into ground stations is limited by the high cost, hardware issues, and complexity of achieving sufficient antenna gain.
%
%
We introduce \textit{ArrayLink}, a distributed phased array architecture that coherently combines multiple small commercially available panels to achieve high‐gain beamforming and unlock line‐of‐sight MIMO spatial multiplexing with minimal additional capital expenditure. By spacing 16 \(32\times32\) panels across a kilometer‐scale aperture, \textit{ArrayLink} enters the radiative near-field, focusing energy in both angle and range while supporting up to four simultaneous spatial streams on a single feeder link.
Through rigorous theoretical analysis, detailed 2D beam pattern simulations and real-world hardware experiments, we show that \textit{ArrayLink} (i) achieves dish-class gain with in range 1-2\,dB of 1.47 m reflector, (ii) maintains four parallel streams at ranges of hundreds of kilometers (falling to two beyond 2000 km), and (iii) exhibits tight agreement across theory, simulation, and experiment with minimal variance. These findings open a practical and scalable path to boosting LEO backhaul capacity. 
\end{abstract}



\section{Introduction}
\label{sec:introduction}

The rapid growth of low Earth orbit (LEO) satellite constellations has fundamentally transformed broadband connectivity, Earth observation, and direct-to-device connectivity. Major players like SpaceX's Starlink, Planet Labs, Amazon Kuiper, and OneWeb \cite{SpaceX_wiki, Planet_Labs_wiki, Project_Kuiper_Wiki, Eutelsat_OneWeb_wiki} are deploying massive networks that underpin critical applications, from internet services to solar weather monitoring. However, the ground segment infrastructure has not scaled proportionally, creating a significant bottleneck in satellite-to-ground backhaul capacity. Overcoming this issue requires ground station architectures capable of delivering (1) \emph{high throughput} to handle increasing data rates, (2) \emph{resource efficiency} to quickly track and seamlessly transfer data between fast moving satellites, and (3) \emph{scalability} to economically deploy numerous ground stations in response to expanding satellite constellations.

\begin{figure}[t]
     \centering
     \begin{subfigure}[t]{0.23\textwidth}
         \centering
\includegraphics[width=\textwidth]{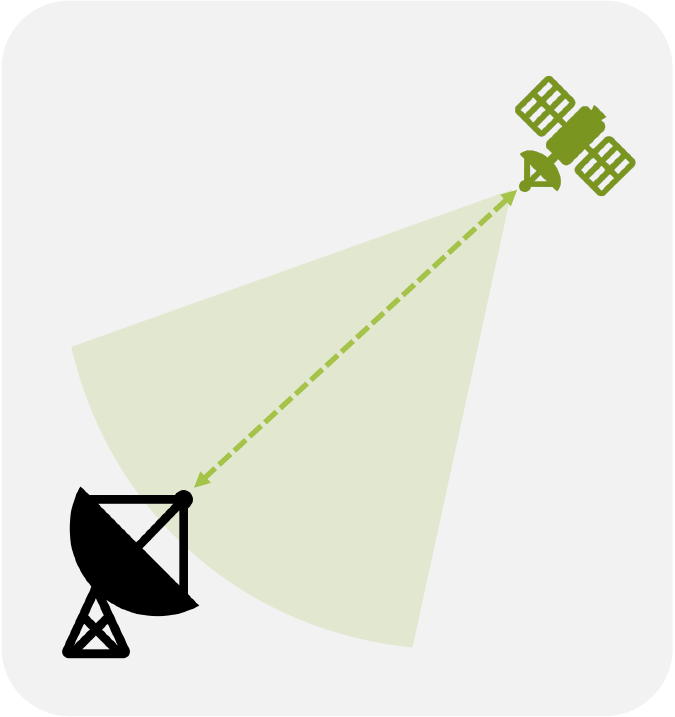}
         \caption{Current satellite ground station built using Parabolic Dishes.}
         \label{fig:intro_parabolic_dish}
     \end{subfigure}
     \hfill
     \begin{subfigure}[t]{0.23\textwidth}
         \centering
         \includegraphics[width=\textwidth]{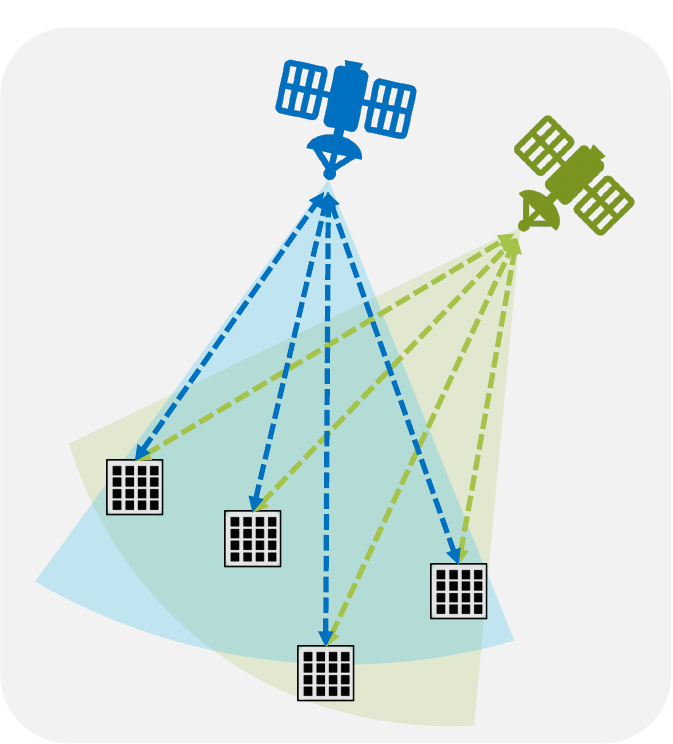}
         \caption{Proposed satellite ground stations using distributed phased arrays.}
         \label{fig:intro_distributed_arrays}
     \end{subfigure}
     \caption{
     An illustration of satellite to earth ground station links: (a) current approach (b) proposed approach}
    \label{fig:motivation_satellite_links}
    \vspace{-0.02\textheight}
\end{figure}

\textbf{Parabolic dish antennas} remain the backbone of satellite ground stations for high‐gain feeder links, providing reliable, focused beams. For example, SpaceX employs 1.47\,m and 1.85\,m dishes, achieving gains of 49.5 and 52.6\,dBi, respectively~\cite{fcc_spacex_LA_gndstn, fcc_spacex_gnd_stn}. However, these dishes are inherently inflexible, as each can only track one satellite at a time. Mechanical steering is required to maintain alignment with rapidly moving LEO satellites, resulting in significant downtime during satellite handoffs. For instance, Intelsat dishes rotate at speeds of only $2 - 5^\circ$ per second, causing transitions from $-60^\circ$ to elevation $+60^\circ$ to take nearly a minute, during which the station is temporarily unavailable. These constraints lead to inefficient resource utilization. Furthermore, deploying additional parabolic dishes to scale capacity significantly increases land, power, and backhaul costs, making traditional dish architectures unsuitable for meeting the flexibility and scalability demands of modern LEO constellations.

\textbf{Large phased arrays} offer a promising alternative, enabling beam hopping in microseconds and supporting multiple links simultaneously without mechanical parts \cite{he2021review}. Despite these advantages, their integration into ground stations faces significant challenges. Achieving a high antenna gain comparable to a 1.85 m Starlink class dish (52.6 dBi \cite{fcc_spacex_gnd_stn}) would \textit{require a large array with more than 50,000 elements}. Such massive monolithic arrays lead to prohibitive power consumption, complex thermal management, and high manufacturing costs, severely limiting their current field deployment.  Consequently, neither traditional parabolic dishes nor large phased-array architectures effectively meet the flexibility, scalability, and performance demands of next-generation LEO ground stations.

\textbf{\sysname:} To overcome these limitations, we introduce \sysname, a novel distributed phased-array ground station that coherently links many small phased-array panels spread across a large area into a single high-performance system. By efficiently combining multiple affordable and commercially available arrays, \sysname~delivers high beamforming gains and simultaneously unlocks multiple concurrent data streams, transforming satellite ground station connectivity. Specifically, \sysname~addresses all three critical requirements: (1) achieves \textit{high link throughput} by delivering high-gain links and allowing multiple streams (2) the inherent electronic beam-steering capability of phased arrays allows rapid satellite tracking and \textit{efficient spectrum utilization}, eliminating downtime associated with mechanical steering; (3) leveraging mass-produced phased-array panels developed originally for user terminals and in-flight connectivity drastically reduces deployment costs and enables rapid scalability.

To achieve high-speed, scalable backhaul connectivity, \sysname~addresses two fundamental challenges:
\begin{itemize}
    \item \textbf{Achieving Cost Effective High Gain Links:}
    A key barrier for phased arrays in ground-station architectures is matching the high gain provided by large dishes; for instance, SpaceX's 1.85\,m dishes achieve 52.6\,dBi~\cite{fcc_spacex_gnd_stn}. Achieving similar gain with phased arrays traditionally requires over 50,000 (very small) antenna elements, making monolithic arrays expensive and complex. Our key observation is that antenna gain increases logarithmically with the number of elements—rapidly rising initially but flattening with larger counts. Rather than building prohibitively large arrays, we combine a modest number of commercially available phased-array panels. For example, by coherently combining sixteen typical user-terminal panels (each with about 36.1\,dBi), \sysname~achieves approximately 48.1\,dBi. To bridge the remaining gap (~4.5\,dB) without inefficiently adding more elements, we leverage digital multiple-input multiple-output (MIMO) techniques to enable multiple simultaneous data streams, maximizing throughput while controlling costs.

    \item \textbf{Enabling Multi-Stream MIMO in LoS Channels:}  
    Conventional satellite links operate primarily in a Line-of-Sight (LoS) environment, leading to highly correlated MIMO channels that are unsuitable for spatial multiplexing. However, In near-field conditions, each antenna experiences distinct phase variations, transforming the channel suitable for MIMO. However, The key challenge here is \textit{"how to enable these near-field conditions at practical satellite distances"}. To address this, we developed a novel mathematical model that precisely characterizes near-field MIMO feasibility, showing explicitly how adjusting transmit and receive aperture sizes can control the near-field region.
    Leveraging this model, we distribute phased-array panels across a kilometer-scale aperture, enabling robust near-field LoS MIMO and supporting multiple concurrent streams to satellites at distances up to 2,000\,km, all while maintaining high individual link gains (48.1\,dBi). 

\end{itemize}

Further, we discuss how reduces grating lobes and achieve coherent combining in our design section. We validate the effectiveness of \sysname through a rigorous evaluation, combining theoretical insights, high-fidelity simulations, and real-world hardware experiments. First, we derive analytical models that precisely characterize singular values, degrees of freedom, and define clear boundaries for near-field MIMO feasibility based on transmit and receive aperture sizes. Next, we perform outdoor hardware experiments with a 2×2 MIMO setup at 27\,GHz, testing various aperture sizes and ranges from 2.5\,m to 100\,m. Remarkably, our hardware measurements closely align with both theoretical predictions and simulation results, demonstrating the accuracy and robustness of our simulator and mathematical models. Finally, extensive satellite-to-ground station link simulations confirm that \sysname coherently focuses beams in both angle and distance, achieving individual link gains of approximately 48.14\,dBi. Notably, our results show the ability to sustain up to four simultaneous data streams at hundreds of kilometers and at least two streams at ranges beyond 2,000\,km—highlighting \sysname's powerful capability to transforming satellite ground-station connectivity.

\textbf{Contributions.} We summarize our contributions as follows:
\begin{enumerate}
  \item \textbf{\sysname architecture:} A novel and scalable ground station design that coherently combines multiple small phased array panels to enable high-gain, multi-stream feeder links.
  \item \textbf{Near-field MIMO modeling:} An analytical framework that characterizes the feasibility of spatial multiplexing in LoS channels and demonstrates how aperture dimensions can be tuned to achieve near-field MIMO at satellite-scale distances.
  \item \textbf{Coherent beamforming:} A practical delay-compensation technique that enables phase-coherent combining across distributed panels while mitigating grating lobes through aperiodic panel placement. 
  \item \textbf{Open-source tools and datasets:} A publicly released Python-based simulator for near-field MIMO along with a real-world hardware dataset (captured at 27\,GHz over 2.5–100\,m), supporting reproducibility and future research in LoS MIMO.
\end{enumerate}

\section{Background and Related Work}\label{sec:background}
We begin by reviewing traditional parabolic dish antennas and the alternative approach of using large phased arrays for satellite ground stations. We then discuss related work on near-field and line-of-sight (LoS) MIMO, highlighting key challenges in deploying large phased arrays for practical ground-station applications.

\subsection{Background on Satellite ground station geometry}
\begin{figure}[t]
    \centering
    \includegraphics[width=0.4\textwidth]{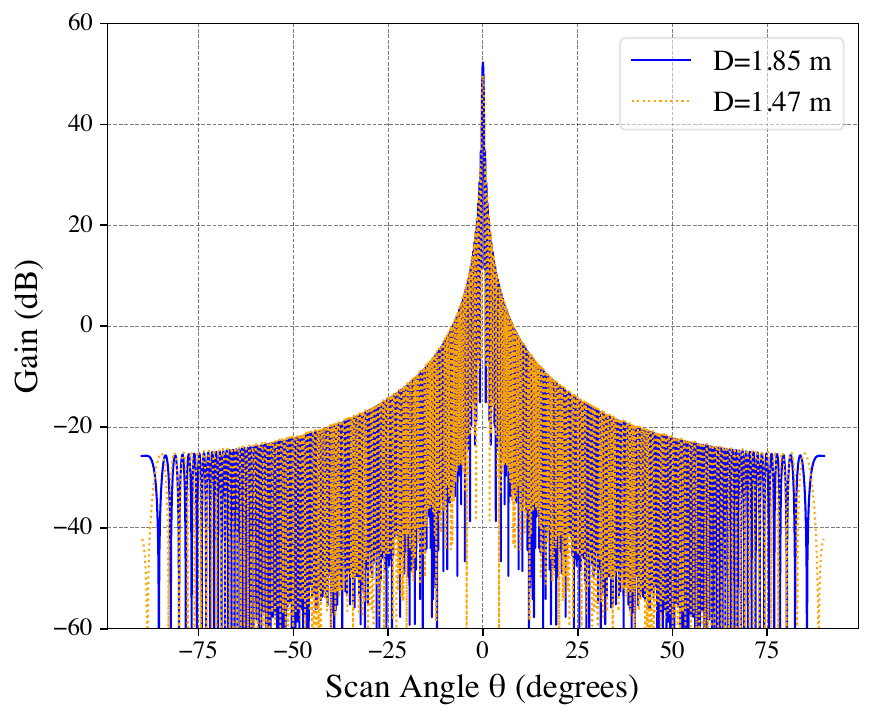}
    \caption{Gain pattern of a 1.85m (max 52.6 dBi) and 1.47m (max 49.5 dBi) parabolic dish antenna operating at 28 GHz.}
    \label{fig:parabolic_beam_pattern}
    \vspace{-0.02\textheight}
\end{figure}

\textbf{\textit{Parabolic Dish Ground Stations: }}
Parabolic dish antennas have been the de facto standard for satellite ground stations due to their high directivity and excellent gain characteristics. The antenna gain \( G \) is given by \cite{balanis2016antenna}:
\[
G = \frac{4\pi A}{\lambda^2}e_A = \left(\frac{\pi D}{\lambda}\right)^2 e_A
\]
where \( D \) is the dish diameter, \( \lambda \) the wavelength, and \( e_A \) the aperture efficiency (typically 0.5–0.7). For example, a 1.47 m parabolic dish operating in the Ka band (28 GHz) achieves approximately 49.5 dBi gain, sufficient to support Starlink LEO constellations \cite{fcc_spacex_LA_gndstn}. As illustrated in Fig.~\ref{fig:parabolic_beam_pattern}, the measured gain pattern closely approaches 48.1 dBi for a 1.47 m dish and 52.6 dBi for a 1.85 m dish at 28 GHz.

However, for LEO satellites traversing the sky at approximately 7.6 km/s, mechanical steering systems must continuously reorient parabolic dishes, introducing latency, handover interruptions, and added operational complexity. Transitions between satellites may require angular movements exceeding 120°, which can take several seconds to a minute depending on the dish size \cite{ortbital_rot_per_sec}. At higher frequencies such as 28 GHz, parabolic dishes also demand extremely precise pointing to maintain link quality and avoid significant signal degradation. Even minor pointing errors at these frequencies can result in substantial reductions in received signal strength and increased interference \cite{weerackody2006motion, geng2021correction}. These constraints lead to periods of unavailability, wasting valuable communication time and leaving system resources underutilized.

\textbf{\textit{Phased Array Antennas: }}
Phased array antennas leverage electronically steerable beams without moving parts. The beam is steered by adjusting the phase of \( N \) radiating elements, achieving a directional gain that depends on both the element pattern and array factor \cite{wiki_phased_array}:]
The array factor \( AF(\theta, \phi) \) for a rectangular \( M \times N \) element array with uniform spacing \( d_x, d_y \) and corresponding weights \( w_{mn}\) is given by:



\begin{equation}
AF(\theta,\phi) = \sum_{m=0}^{M-1} \sum_{n=0}^{N-1} w_{mn} e^{jk[md_x \sin(\theta) \cos(\phi) + nd_y \sin(\theta) \sin(\phi)]}
\label{eq:array_factor}
\end{equation}

where \( k = 2\pi/\lambda \) is the wave number and \( \lambda \) is the wavelength.

These antennas can form multiple independent beams, enabling simultaneous tracking of several satellites and steering less than a millisecond rather than seconds \cite{wiki_phased_array}. However, scaling these systems for ground station applications remains a challenge. Achieving gains exceeding 50 dBi would require tens of thousands of antenna elements (e.g., 20,000+ elements at approximately 6 dBi each), resulting in substantial power consumption, thermal management complexity, mutual coupling effects, and high manufacturing costs.

In summary, parabolic dishes provide high gain but limited agility, while phased arrays offer agility at the expense of increased complexity and cost. Neither approach alone satisfies the scalability and performance demands of next-generation LEO ground stations.

\subsection{Related work}


\textbf{Line-of-Sight MIMO:} 
Recent work on near-field wireless studies on beam focusing for extremely-large aperture arrays \cite{kang2025nmap, huang2025symmetric} and channel estimation \cite{li2025blind, liu2025gradient, lei2025near, zhou2025super} investigate how spherical-wavefronts can be exploited with a variety of analog/digital beamformers \cite{liu2025dynamic}.
More recent work focuses on reconfigurable intelligent surfaces \cite{zhou2025ris} and integrated sensing-and-communication (ISAC) \cite{sun2025near} to extend the near-field model to new hardware or dual-use scenarios.
Recent surveys \cite{cui2022near, parvini2025tutorial} clarify Rayleigh/Fresnel distance definitions and introduce Fresnel-zone beam-focusing concepts. None of these works, however, examine how element-placement strategies influence capacity or sidelobe behaviour when hardware budgets constrain the number of antennas.

Classical LoS MIMO studies take the complementary view of improving rank through geometry. Sarris et al. \cite{sarris2005maximum} and Jiang et al. \cite{jiang2022design} show that carefully spaced, uniform planar arrays can achieve full multiplexing gain in far-field LoS channels. Unfortunately, scaling their prescriptions to a satellite ground station would demand thousands of elements, making cost, power, and grating-lobe suppression prohibitive.
At the network level, \cite{fernandez2024constellation} raises channel rank by distributing LEO satellites across multiple orbital planes, but ground-station complexity remains unchanged.
To the best of our knowledge, no prior work applies near-field LoS-MIMO principles to the design of LEO satellite ground stations. Our study fills this gap by proposing a randomized, sparse-aperture phased-array whose per-element phase programming simultaneously preserves near-field multiplexing gains and suppresses aliasing-induced side lobes, achieving high capacity with an order-of-magnitude fewer antennas than uniform designs.

\textbf{Phased-array ground stations.}
Several recent efforts aim to replace bulky satellite dishes with electronically steerable phased arrays.
\cite{pan2023pmsat} introduces a ground terminal that combines a small 1x4 phased array with a passive 21x21 metasurface to enhance signal strength. While effective in boosting link budget, this design operates in a far-field, single-beam mode and does not leverage MIMO gains or explore non-uniform antenna placement.
\cite{chang2024smart} presents a “smart transfer planer” that uses three 8×8 arrays spaced at half-wavelength and connected via a compact Rotman lens to enhance indoor satellite reception. However, this setup still assumes uniform spacing and far-field beam steering, which limits its ability to suppress sidelobes or support spatial multiplexing.
\cite{merino2022phased} Performs electromagnetic simulations of a large planar array with 8,910 elements covering 2 m$^2$, designed for Starlink-class gateways. While the study provides a detailed link budget analysis, it relies on a fully populated array and does not consider reducing antenna count or exploiting near-field effects.
\cite{adomnitei2024phased} Surveys AI-assisted beam tracking and software-defined radio (SDR) techniques for LEO satellite terminals, but does not address near-field propagation or array geometry design. These works primarily focus on gain enhancement or tracking in the far-field regime. In contrast, our approach applies near-field LoS-MIMO principles to the design of sparse ground station arrays. We show that by randomizing antenna placement and optimizing phase vectors, it is possible to significantly reduce antenna count while maintaining capacity and controlling side lobe levels.

\begin{figure}[!t]
     \centering
     \begin{subfigure}[t]{0.23\textwidth}
         \centering
         \includegraphics[width=\textwidth]{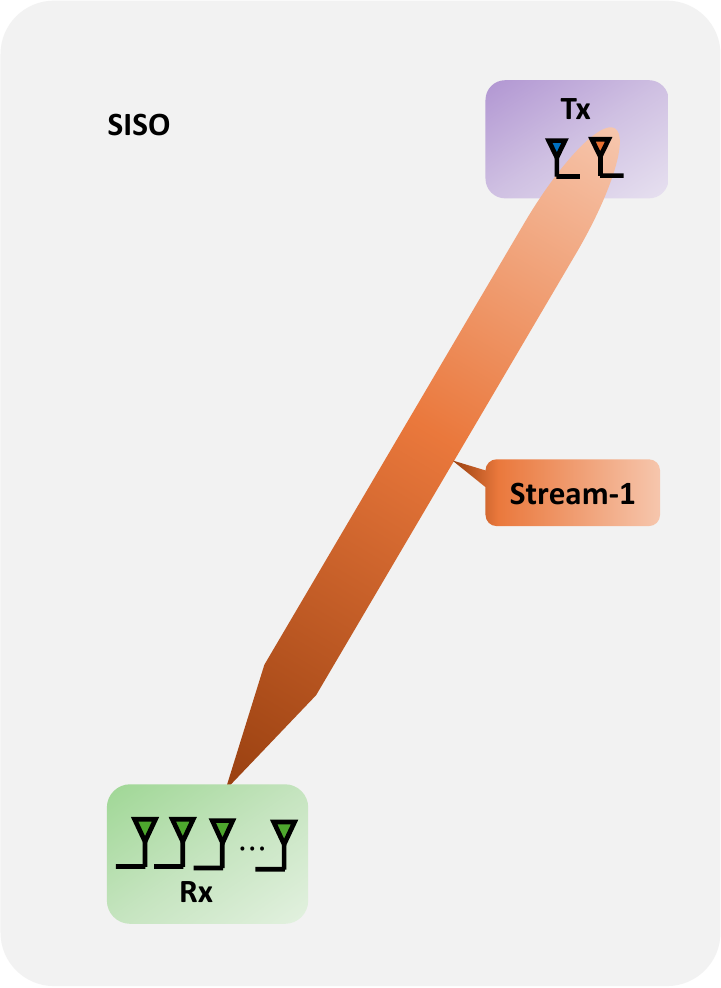}
         \caption{Uniform linear phased arrays (ULA/UPA) only allows SISO (Single stream) satellite link.}
         \label{fig:design_siso_scenario}
     \end{subfigure}
     \hfill
     \begin{subfigure}[t]{0.23\textwidth}
         \centering
         \includegraphics[width=\textwidth]{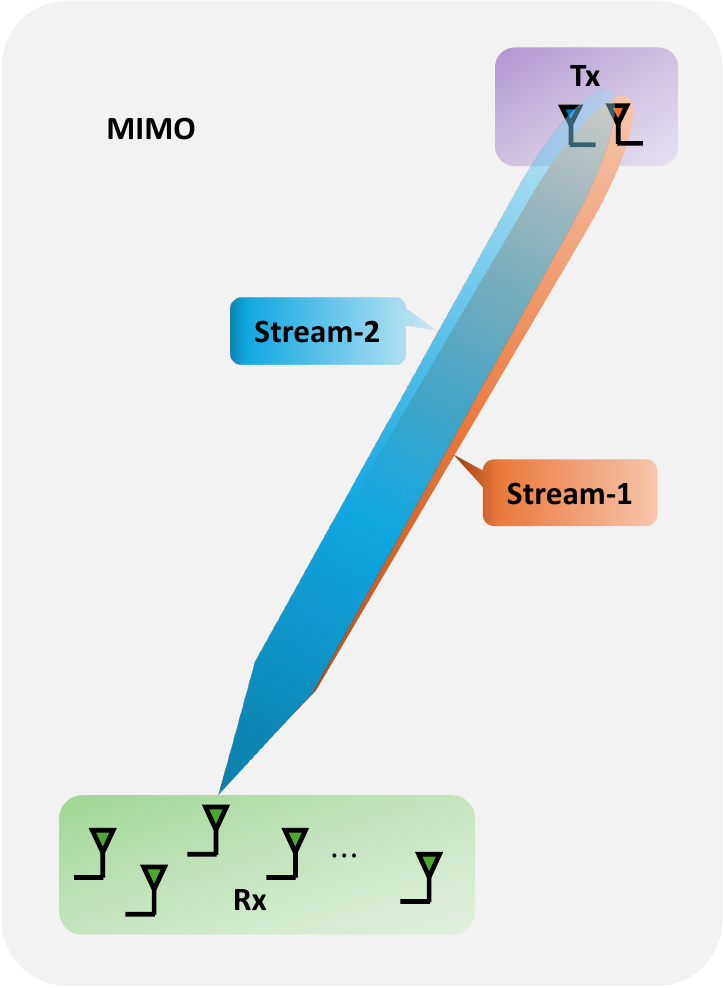}
         \caption{Distributed arrays system enables MIMO (Multiple simultaneous streams) satellite links.}
         \label{fig:design_mimo_scenario}
     \end{subfigure}
     \caption{
     An illustration of satellite to earth ground station links: (a) Large (ULA/UPA) phased arrays (b) Proposed distributed phased arrays approach. 
     }
    \label{fig:design_mimo_motivation}
    \vspace{-0.02\textheight}
\end{figure}





\section{Design }\label{sec:design}
To overcome the limitations of both parabolic dishes and large phased arrays for ground station, we propose a distributed smaller phased array-based architecture (\sysname). As illustrated in Fig.~\ref{fig:intro_distributed_arrays}, the system aggregates multiple smaller phased array panels into a coordinated network. The design aims to (i) achieve high-throughput backhaul links, (ii) support multiple simultaneous spatial streams through MIMO, and (iii) ensure practical feasibility and cost-effectiveness for real-world deployment.

\subsection{Aggregating Smaller Phased Arrays for High-Gain Links}
\label{subsec:combining_phased_arrays}

Achieving high-gain satellite links traditionally relies on parabolic dishes, whose large physical surface areas naturally yield high effective apertures and concentrated energy beams. In contrast, phased arrays provide dynamic beam steering but individually offer moderate gain due to practical limitations in array size and element counts \cite{balanis2016antenna}. To overcome this constraint, \sysname adopts an architecture that aggregates multiple smaller phased array units, each contributing modest gain, into a distributed but coherently combined system. This approach leverages the scalability and flexibility advantages inherent in phased array technology.

In \sysname, each ground station comprises \(N\) phased array panels, each containing \(M \times M\) antenna elements. These panels are geographically dispersed within a specified total aperture \(A_\mathrm{total}\), enabling strategic placement to minimize grating lobes and enhance spatial resolution. By coherently combining signals from all distributed panels, the total achievable array gain \(G_\mathrm{total}\) can be expressed as:
\begin{align}
G_\mathrm{total} &= 10\log_{10}\bigl(N \cdot G_{\mathrm{PA}} \cdot e^{-\delta}\bigr)\ \mathrm{dBi}, \nonumber\\
&\approx 10\log_{10}(N) + G_{\mathrm{PA}}\ \mathrm{(dB)}, \label{eq:N_Gpa}
\end{align}
where \(G_\mathrm{PA}\) represents the gain of a single phased array panel, and \(\delta\) accounts for phase misalignments due to synchronization errors across distributed panels.

Commercial satellite terminals from Starlink, Kuiper, OneWeb, and Telesat have already demonstrated planar phased arrays in delivering dynamic, low-latency satellite access \cite{fcc_spacex_phased_arrays, kuiper_terminals, fcc_oneweb_phased_arrays}. \textbf{\sysname aggregates these smaller and commercially available phased arrays to achieve higher gains} required for robust ground station backhaul links. The essential question then becomes \textit{`How many of these phased arrays required?'}. For instance, current state-of-the-art phased arrays feature approximately 1,024 elements (e.g., \(32 \times 32\)), each with typical microstrip antenna element gains around $\approx6$ dBi \cite{extereme_waves_pa, wiki_microstrip_antenna}. 
Such a panel achieves a total gain of roughly 36.1 dBi, comprising 30.1 dB of array gain plus the element gain. Consequently, based on Equation~\ref{eq:N_Gpa}, achieving a gain comparable to Starlink’s standard 1.47 m parabolic dish (~48.1 dBi) \cite{fcc_spacex_LA_gndstn} requires aggregation of approximately 16 such phased array panels.

\begin{figure}[!t]
     \centering
     \begin{subfigure}[t]{0.23\textwidth}
         \centering
         \includegraphics[height=0.21\textheight]{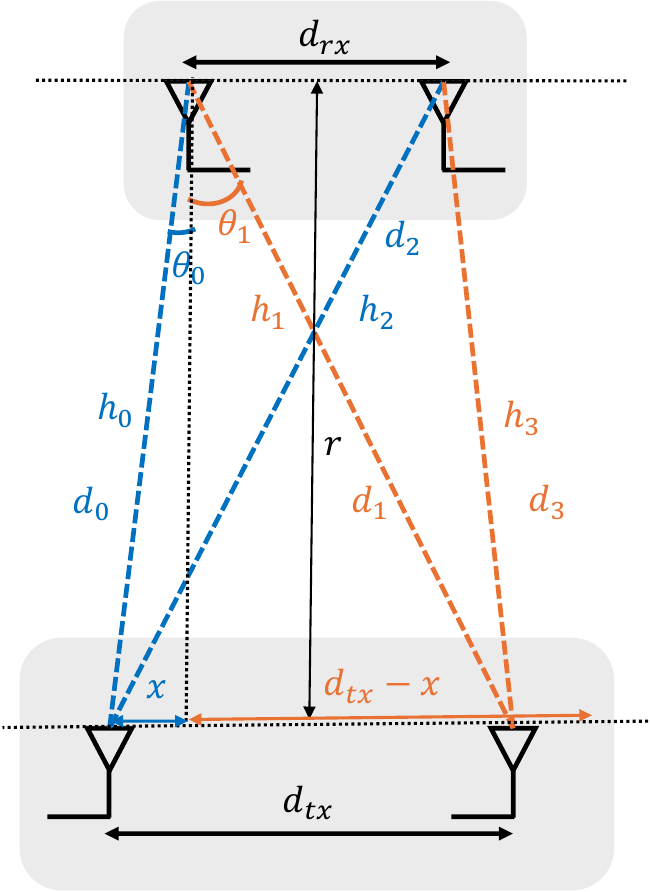}
         \caption{2x2 MIMO Scenario with transmitter and receiver perpendicular to LoS path.}
         \label{fig:experimental_setup}
     \end{subfigure}
     \hspace{0.005\textwidth}
     \begin{subfigure}[t]{0.23\textwidth}
         \centering
         \includegraphics[height=0.21\textheight]{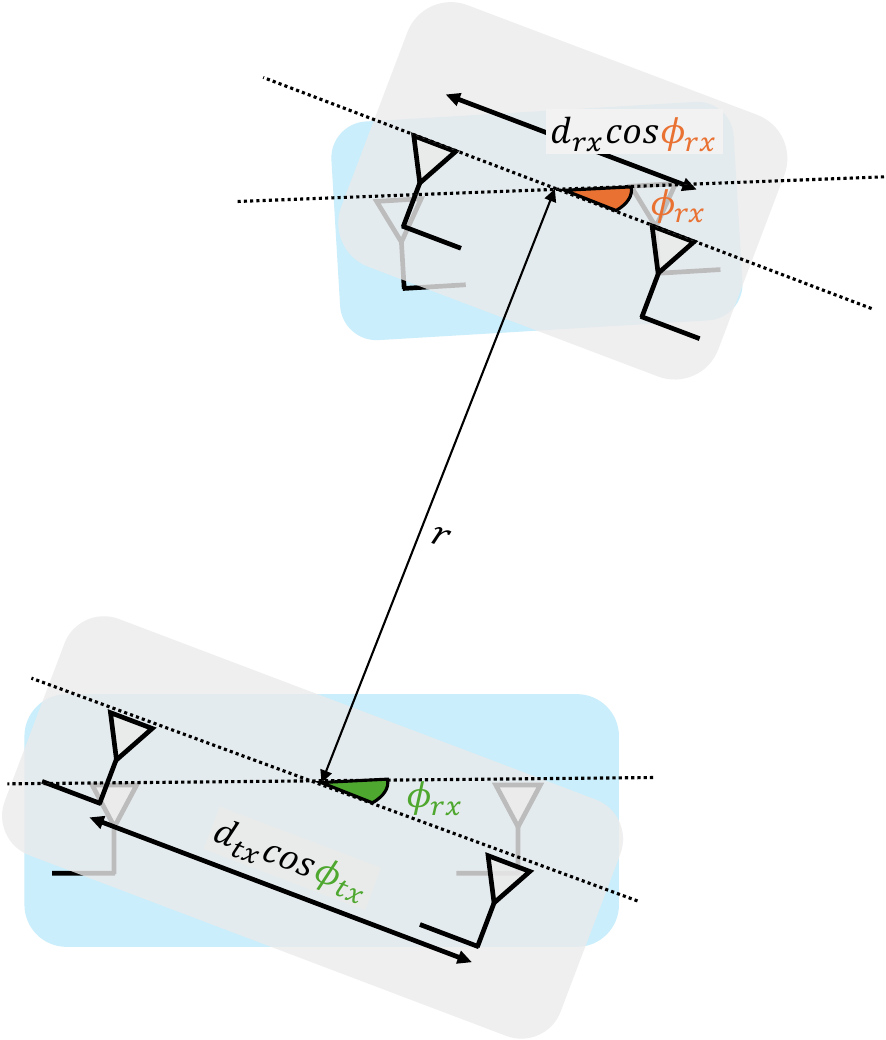}
         \caption{Virtual aperture sizes for 2x2 MIMO scenario with transmitter and receiver at different angles to LoS path.}
         \label{fig:experimental_setup_angle_deviation}
     \end{subfigure}
     \caption{
     Line-of-Sight (LoS) MIMO Scenario with two transmit antennas and two receiver antennas. 
     }
    \label{fig:design_setup}
    \vspace{-0.02\textheight}
\end{figure}

Despite these advantages, a critical limitation remains: phased arrays inherently experience an effective aperture reduction as the satellite moves away from the nadir direction. Specifically, aperture effectiveness declines proportionally to \(\sin(\theta)\), where \(\theta\) is the off-nadir angle. Practically, maintaining high-gain links across wide elevation angles (e.g., 30° to 150°, spanning ±60° from nadir) results in significant gain reductions - up to 6 dB losses at the coverage edges. A straightforward, albeit impractical, approach to compensate for this loss is to quadruple the number of arrays (from 16 to 64 panels), significantly increasing cost, power, and operational complexity \cite{mailloux2017phased, hansen2009phased}.

Thus, while aggregating smaller phased arrays offers a promising strategy for scalable high-gain links, simply scaling the number of arrays proves insufficient in practice. \sysname addresses this fundamental limitation by harnessing near-field MIMO techniques to unlock multiple concurrent spatial streams, ensuring sustained high throughput even as individual array gains fluctuate due to aperture variations.

\subsection{Near-Field MIMO: Unlocking Spatial Streams}
\label{sec:design:nearfield_mimo}

The key idea in this section is to investigate whether it is possible to maintain high throughput not by relying on a single high-SNR link, but by utilizing multiple spatial streams with lower individual SNRs. For example, instead of transmitting 6 bits per symbol using 64-QAM, one could transmit two parallel streams, each using lower-order modulation (e.g., 4–5 bits/symbol), and achieve comparable or higher aggregate throughput. This raises a central question: \textit{how can we enable MIMO or support multiple spatial streams in satellite-to-ground feeder links where the channel is predominantly Line-of-Sight (LoS) and lacks rich scattering?}.

In conventional satellite links, the LoS-dominated channel leads to highly correlated entries in the MIMO channel matrix, rendering it ill-conditioned and rank-deficient. To assess feasibility, we consider a simplified 2$\times$2 system where both the transmitter and receiver employ two antennas, as shown in Fig.~\ref{fig:experimental_setup}. 
Let the MIMO channel matrix be:
\[
H = 
\begin{pmatrix}
 h_0 & h_1 \\
 h_2 & h_3
\end{pmatrix}
\]
where each $h_i$ denotes the complex baseband channel between a transmit-receive antenna pair and is modeled as:
\begin{equation}
    h_i = \frac{\lambda}{\sqrt{4\pi d_i^2}} e^{-j \frac{2\pi d_i}{\lambda}}
    \label{eq:h_eq}
\end{equation}
with $d_i$ representing the path length and $\lambda$ the carrier wavelength. 

As the distance between the transmitter and receiver exceeds the Fresnel distance, given by $r_\mathrm{Fresnel} = 0.62 \sqrt{\frac{d^3}{\lambda}}$, where $d$ is the maximum aperture size and $\lambda$ is the carrier wavelength, the channel enters the radiative near-field regime~\cite{balanis2016antenna}. With a further increase in distance beyond the Fraunhofer distance, $r_\mathrm{Far} = \frac{2d^2}{\lambda}$, the channel transitions to the far-field, where planar wavefront approximations become valid~\cite{balanis2016antenna}. In both the radiative near-field and far-field regimes, the amplitude variation across antennas is typically negligible, and phase differences dominate the channel characteristics \cite{liu2023near}. Therefore, we normalize the amplitude and retain only the phase component in our channel model:
\[
h_i = e^{-j \theta_i}, \quad \text{where} \quad \theta_i = \frac{2\pi d_i}{\lambda}, \quad |h_i| = 1.
\]

For such a unit-modulus complex matrix, the singular values $\sigma_1, \sigma_2$ of $H$ are determined by the phase spread (derivation given in appendix):
\begin{equation}
    \sigma_{1,2} = \sqrt{2 \pm 2\left|\cos \frac{\Delta}{2} \right|},
    \label{eq:sigma_eq}
\end{equation}
where the phase spread $\Delta$ is given by
\[
\Delta = (\theta_0 + \theta_3) - (\theta_1 + \theta_2).
\]

For our 2x2 setup (Fig.\ref{fig:experimental_setup}) assuming LoS propagation, this expression simplifies to:
\begin{align}
\Delta &= \frac{2\pi}{\lambda} \left[ -(d_0 - d_2) + (d_1 - d_3) \right] \nonumber\\
&= \frac{2\pi}{\lambda} d_\mathrm{tx} \left[ -\sin(-\theta_0) + \sin(\theta_1) \right] \nonumber\\
&= \frac{2\pi}{\lambda} d_\mathrm{tx} \left( \frac{x}{d_0} + \frac{d_tx - x}{d_1} \right) \nonumber\\
& \approx \frac{2\pi}{\lambda} d_\mathrm{tx} \left( \frac{x}{r} + \frac{d_\mathrm{rx} - x}{r} \right) \nonumber\\
\Delta &= {2\pi} \cdot \frac{d_\mathrm{tx} d_\mathrm{rx}}{\lambda r}
\label{eq:delta_standard}
\end{align}

where $d_\mathrm{tx}$ and $d_\mathrm{rx}$ are the transmission and reception antenna spacings, and $r$ is the distance between center of transmitter to receiver.

\begin{figure}[!t]
     \centering
     \begin{subfigure}[t]{0.24\textwidth}
         \centering
         \includegraphics[width=\textwidth]{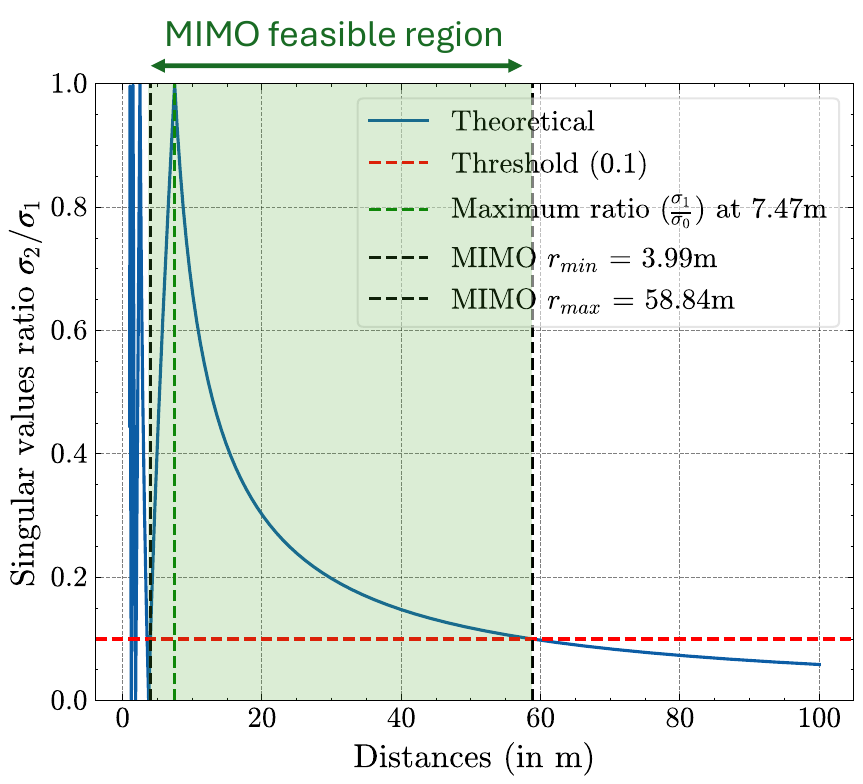}
         \caption{$d_\mathrm{rx}=20 cm$, $d_\mathrm{tx} = 20 cm$}
         \label{fig:design_singular_values_ratio_with_distances}
     \end{subfigure}
     \hfill
     \begin{subfigure}[t]{0.24\textwidth}
         \centering
         \includegraphics[width=\textwidth]{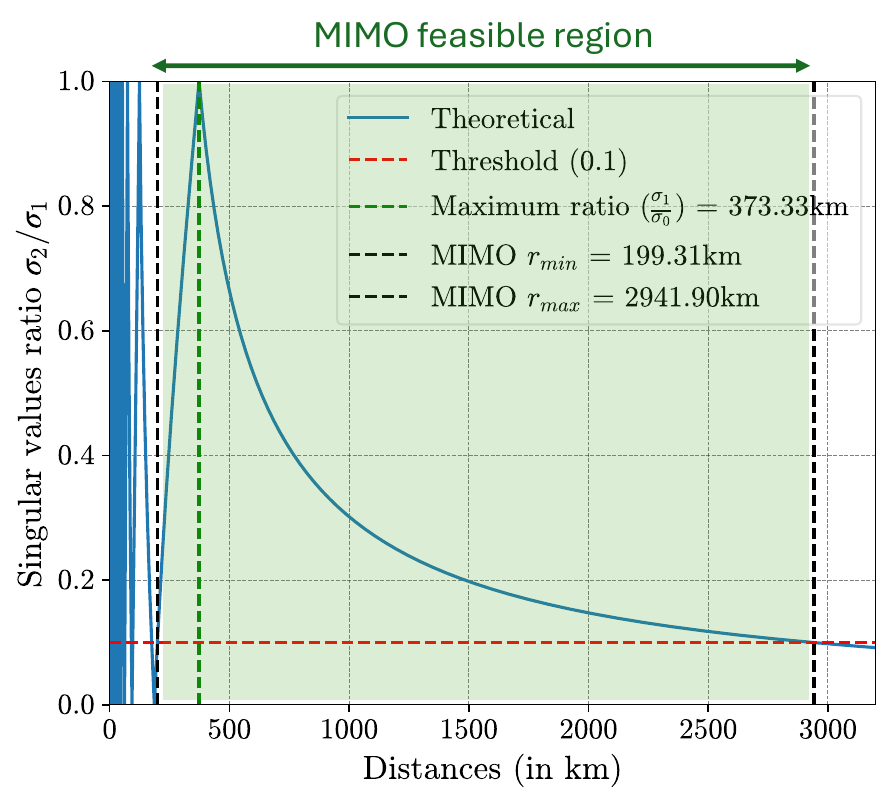}
         \caption{$d_\mathrm{rx}$=$\sqrt{2}km$, $d_\mathrm{tx}$=$\sqrt{2}m$}
         \label{fig:design_singular_values_ratio_with_distances_sat}
     \end{subfigure}
     \caption{
     Illustrating stable min and max distances (boundaries) for MIMO (2x2 system) feasibility: (a) small distances (b) satellite (large) distances.
     }
    \label{fig:design_theoretical_eq_plots}
    \vspace{-0.02\textheight}
\end{figure}

In general, the angles of departure at the transmitter and arrival at the receiver differ from $90^\circ$. As illustrated in Fig.~\ref{fig:experimental_setup_angle_deviation}, the effective inter-element spacings vary with the departure angle $\phi_\mathrm{tx}$ and arrival angle $\phi_\mathrm{rx}$. Consequently, the generalized expression for the phase spread $\Delta$ becomes:
\begin{align}
    \Delta &= 2\pi \frac{ d_\mathrm{tx} \cos(\phi_\mathrm{tx}) \; d_\mathrm{rx} \cos(\phi_\mathrm{rx}) }{\lambda \, r}\nonumber\\
    \Delta &=  2\pi \frac{ \tilde{d}_\mathrm{tx}\;\tilde{d}_\mathrm{rx}}{\lambda \, r} 
\end{align}
where we define the effective spacings $\tilde{d}_\mathrm{tx} = d_\mathrm{tx}\cos(\phi_\mathrm{tx})$, $\tilde{d}_\mathrm{rx} = d_\mathrm{rx}\cos(\phi_\mathrm{rx})$.
For the remainder of this section, we focus on the scenario in Fig.~\ref{fig:experimental_setup}, noting that other angular configurations can be treated by substituting these virtual spacings. In conclusion, this formulation shows that the spacing between the transmit and receive antennas directly determines the phase spread, which in turn governs MIMO feasibility.
In conclusion, this formulation shows that the spacing between tx and rx antennas is correlated to the phase spread, which is determines whether MIMO is feasible.

\begin{figure*}[!t]
     \centering
     \begin{subfigure}[t]{0.32\textwidth}
         \centering
         \includegraphics[width=\textwidth]{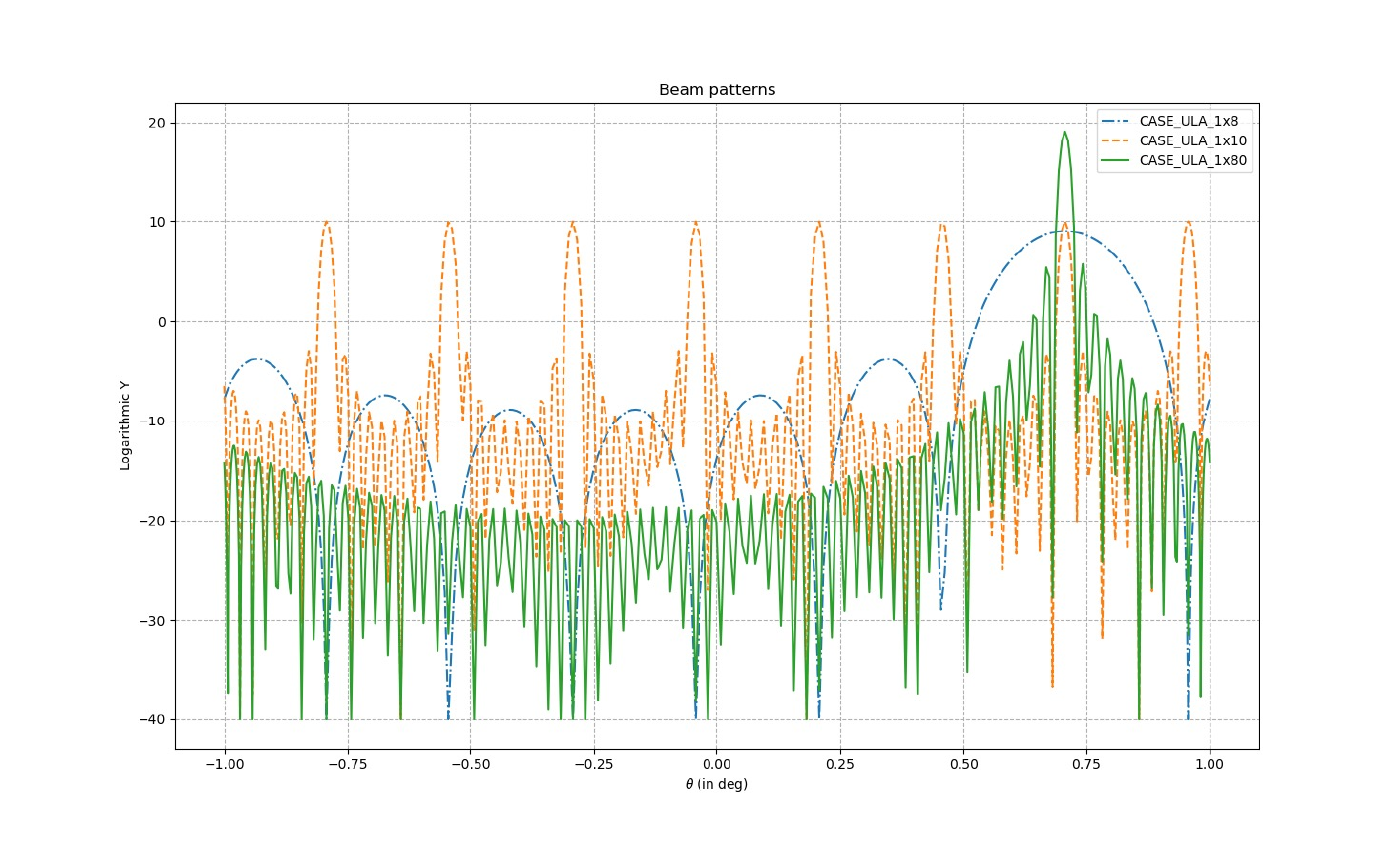}
         \caption{With $\lambda/2$ separation between phased arrays}
         \label{fig:design_ula_combining}
     \end{subfigure}
     \hfill
     \begin{subfigure}[t]{0.32\textwidth}
         \centering
         \includegraphics[width=\textwidth]{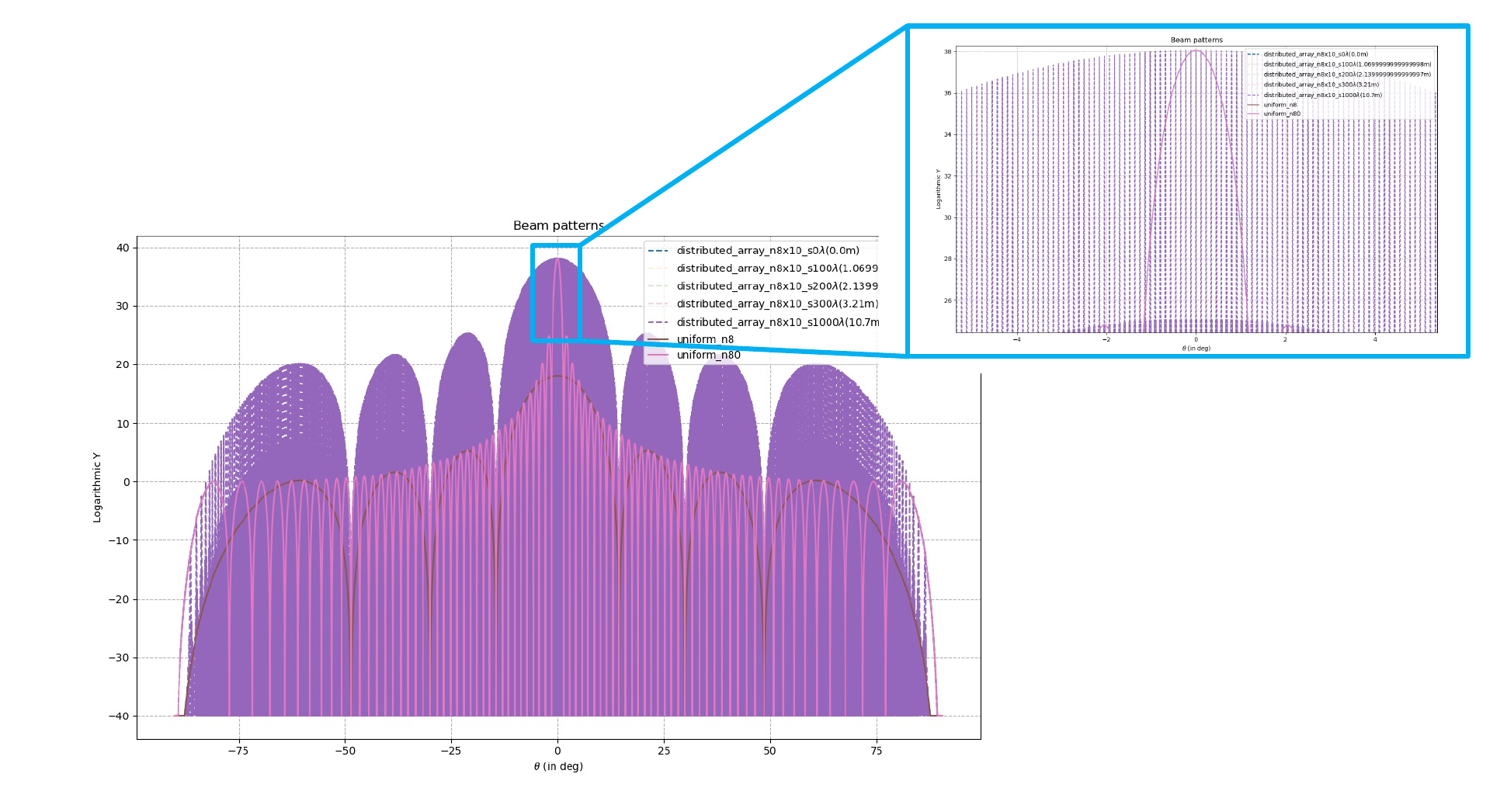}
         \caption{Uniformly spreading phased arrays within $1km$ aperture}
         \label{fig:design_uniform_combining}
     \end{subfigure}
     \hfill
     \begin{subfigure}[t]{0.32\textwidth}
         \centering
         \includegraphics[width=\textwidth]{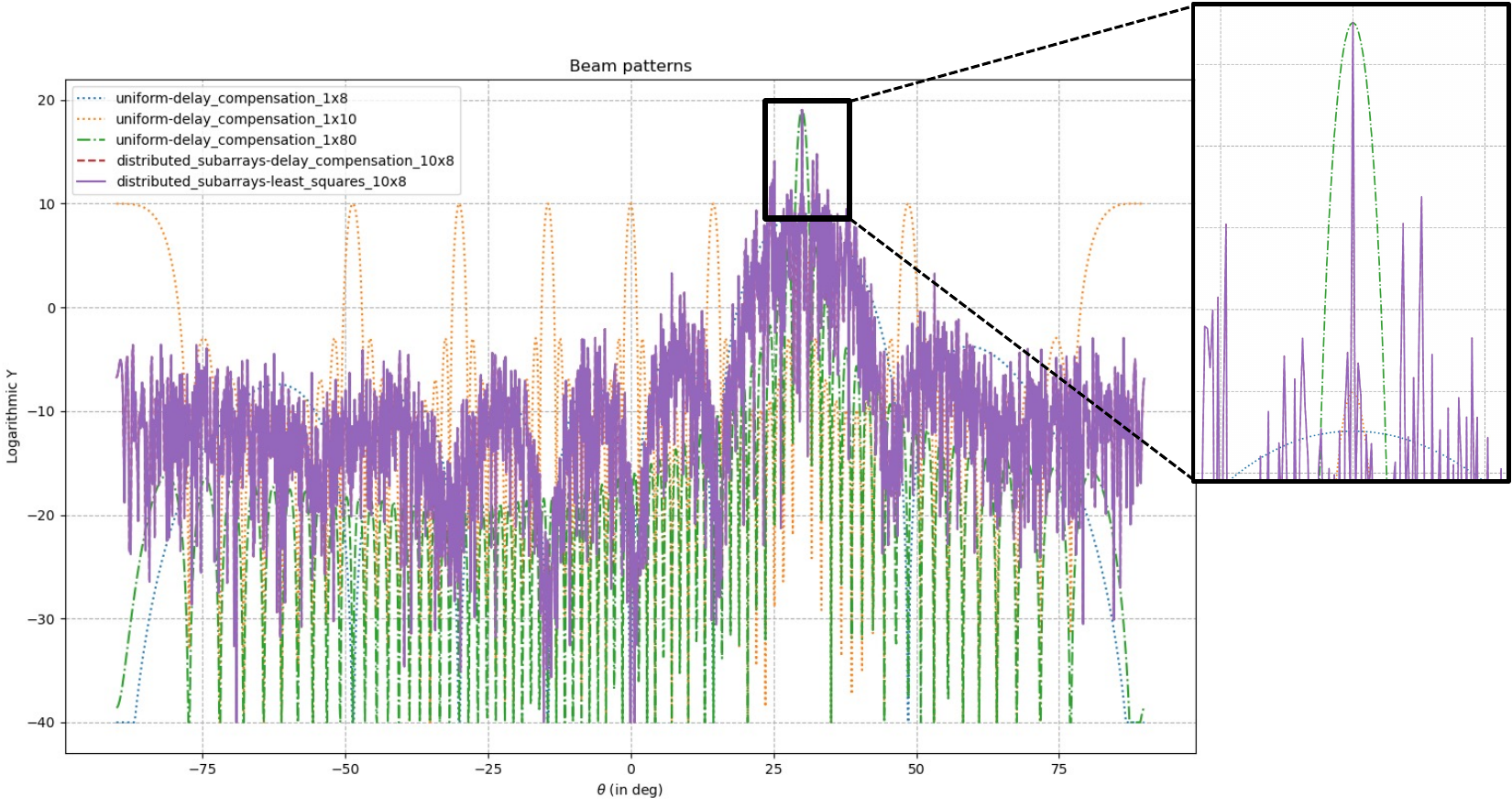}
         \caption{\sysname coherent combining spreading randomly within $1km$}
         \label{fig:design_coherent_combining}
     \end{subfigure}
     \caption{
     Illustrating beam patterns with same number of antenna elements comparing (a) no separation with (b) uniform and (c) random approach for antenna placement. 
     }
    \label{fig:design_combining}
    \vspace{-0.01\textheight}
\end{figure*}

\subsection{Boundaries for MIMO region: min and max distance for a given antenna spacing }
Now that we know MIMO works with multiple spatial streams even in LoS scenarios, the next question is: What defines the boundaries of this MIMO region? When can we reliably enable MIMO? Multipath channel ranks shift with the surroundings, whereas a LoS MIMO channel rank is highly predictable once the antenna spacing and the geometry between transmitter and receiver are fixed.

Ideally, the singular values of $H$ should be nearly equal to maximize MIMO efficiency. 
From Eq.~\eqref{eq:sigma_eq}, equality occurs when $\cos\frac{\Delta}{2}=0$, i.e., $\Delta=\pi$. In practice, the antenna spacings $d_\mathrm{tx}$ and $d_\mathrm{rx}$ are fixed, so the singular‐value ratio varies with range $r$ as follows (illustrated in Fig.\ref{fig:design_theoretical_eq_plots}):
\begin{itemize}
  \item \textbf{Region-1 $\left(r < \dfrac{d_\mathrm{tx}\,d_\mathrm{rx}}{\lambda} \right)$:} Here $\dfrac{\Delta}{2} > \pi  $, causing rapid fluctuations in the singular‐value ratio between 0 and 1 for small changes in $r$.
  \item \textbf{Region-2 $\left(\dfrac{d_\mathrm{tx}\,d_\mathrm{rx}}{\lambda} \leq r \leq \dfrac{2\,d_\mathrm{tx}\,d_\mathrm{rx}}{\lambda} \right)$:}  for $\dfrac{\Delta}{2} = \pi$, yielding $\sigma_2 = 0$.  for $\dfrac{\Delta}{2}$ from $\pi$ to $\pi/2$, changes ratio from $0$ to $1$ and finally at  $\dfrac{\Delta}{2} = \dfrac{\pi}{2}$ gives ratio 1.
  \item \textbf{Region-3 $\left(r > \dfrac{2\,d_\mathrm{tx}\,d_\mathrm{rx}}{\lambda} \right)$:} gives $\dfrac{\Delta}{2} < \dfrac{\pi}{2}$, and ratio gradually decays toward zero. 
\end{itemize}

A key question to address is: What is the minimum and maximum distance that guarantees MIMO? To assess whether MIMO is feasible in a $2 \times 2$ system, a widely used practical condition is based on the ratio of the minimum to maximum singular values of the channel matrix $\mathbf{H}$ \cite{Rohde_Schwarz_MIMO}. If ratio of singular values (condition number) is greater than threshold ($\tau \approx 0.1$) the channel is considered sufficiently well-conditioned to support spatial multiplexing. i.e the condition number for MIMO feasibility is given by
\begin{equation}    
\frac{\sigma_{\min}}{\sigma_{\max}} > \tau (\approx 0.1)
\label{eq:condition_number}
\end{equation}
This criterion reflects the fact that highly unequal singular values lead to poor separation between data streams and degraded performance. Although not a sufficient condition, it provides a reliable heuristic under moderate to high SNRs, aligning with industry assessments that relate low singular value spread to effective MIMO operation.

\textbf{\textit{Min distance:}} 
Establishing a definitive minimum distance is challenging. In Region 1 the condition number fluctuates between 0 and 1. Consequently, even small geometric variations can shift the channel matrix from well conditioned to ill conditioned, resulting in a highly sensitive system. In Region 2, the condition number increases from 0 to 1 and in Region 3 it gradually decays to back to 0, thus yielding a more stable performance criterion. In Region 2, substituting ${1 - cos{\theta}  = 2 sin ^2 {\frac{\theta}{2}}}$ and ${1 + cos{\theta}  = 2cos ^2 {\frac{\theta}{2}}} $. We can further reduce equation (\ref{eq:sigma_eq}) as:
\begin{align}
    \sigma_{max} = sin\frac{\Delta}{4},
    \sigma_{min} = cos\frac{\Delta}{4},
    \frac{\sigma_{min}}{\sigma_{max}} = cot\frac{\Delta}{4} = \frac{1}{tan\frac{\Delta}{4}}
\label{eq:sigma_region2}
\end{align}
Combining equations \ref{eq:delta_standard}, \ref{eq:condition_number} and \ref{eq:sigma_region2}, we get the minimum stable distance to guarantee MIMO as
\begin{equation}
    r_{min} = \frac{\pi}{2\arctan{\frac{1}{\tau}}} \frac{d_\mathrm{tx} d_\mathrm{rx}}{\lambda}
    \label{eq:r_min}
\end{equation}

\textbf{\textit{Max distance:}} In Region 3, as the distance increases, the condition number gradually reduces below the set threshold, making MIMO infeasible. As $\dfrac{\Delta}{4} < \dfrac{\pi}{4}$,  the maximum and minimum singular values switch, i.e.
\begin{align}
    \sigma_{max} = cos\frac{\Delta}{4},
    \sigma_{min} = sin\frac{\Delta}{4},
    \frac{\sigma_{min}}{\sigma_{max}} = tan\frac{\Delta}{4}
\label{eq:sigma_region3}
\end{align}
Similar to calculating the minimum distance, combining  \ref{eq:delta_standard}, \ref{eq:condition_number} and \ref{eq:sigma_region2}, we get maximum distance for MIMO feasibility as follows:
\begin{equation}
    r_{max} = \frac{\pi}{2\arctan{\tau}} \frac{d_\mathrm{tx} d_\mathrm{rx}}{\lambda} \approx \frac{\pi}{2{\tau}} \frac{d_\mathrm{tx} d_\mathrm{rx}}{\lambda}
    \label{eq:r_max}
\end{equation}


Therefore, for distance in range $r_{min} <= r <= r_{max}$, we can determine that two streams are feasible for a 2x2 LoS MIMO system.

\noindent\textbf{Examples.}  
\begin{itemize}
  \item With $d_\mathrm{tx}=d_\mathrm{rx}=0.2\,\mathrm{m}$, $\lambda=0.01\,\mathrm{m}$, and $\tau=0.1$, this yields $r\lesssim62\,\mathrm{m}$, matching our hardware results in Sec.~\ref{subsec:eval_hardware}.
  \item With $d_\mathrm{tx}=2\,\mathrm{km}$, $d_\mathrm{rx}=1\,\mathrm{m}$, and the same $\lambda$ and $\tau$, MIMO remains feasible up to $r\approx2500\,\mathrm{km}$, which is a typical supported range for satellite-ground station feeder links.
\end{itemize}

Further in Sec.~\ref{subsec:eval_simulation_framework}, we demonstrate that distributing 16 phased array panels across a 1.414\,km \(\times\) 1\,km aperture (Figure~\ref{fig:eval_antenna_placement_DSA}) enables multiple simultaneous spatial streams while achieving gains comparable to a parabolic dish. By appropriately positioning the transmit and receive phased arrays, \sysname controls the near-field region to unlock multiple spatial degrees of freedom. This analytical framework establishes the feasibility of LoS MIMO in satellite feeder links without relying on scattering for multipath diversity.

\subsection{Phased Array Placement for Grating-Lobe‐Free Beamforming}

In the previous sections, we derived the spatial separations required to enable multiple simultaneous streams in the feeder link: distributing our phased array panels over a 1 km aperture unlocks LoS MIMO capacity.  However, classical array theory tells us that any uniform spacing beyond $\frac{\lambda}{2}$ introduces grating lobes, making coherent beamforming across km-scale separations seemingly infeasible. How, then, do we coherently enable beamforming while the phased arrays are separated by distances in the order of $10^5\lambda$ ($1\,km$)?

Our key insight is that, for a distributed architecture, the overall beamforming gain factors into two parts: 
(1). the gain of each individual phased array panel, and
(2). the array factor of a virtual “single-element” array whose elements are the panels themselves.   
For example, ten 1×8 linear arrays spaced by $4\lambda$ act equivalently to a single 1×80 uniform linear array (ULA).  Mathematically, the combined beam pattern is the product of the 1×8 ULA pattern (blue) and the 1×10 panel‐placement pattern (orange), which produces the 1×80 ULA pattern (green), as illustrated in Fig.~\ref{fig:design_ula_combining}.

A naive approach would be to place ten 1×8 arrays uniformly with $10\,000\,\lambda$ spacing.  In that case, the panel‐placement pattern becomes a finely sampled impulse train (Fig.~\ref{fig:design_uniform_combining}), and when multiplied by the 1×8 ULA pattern yields rapid fluctuations in gain: even a $0.0001^\circ$ steering error leads to a deep null.  Such a beam pattern is highly unstable and impractical to use efficiently.

\sysname instead adopts a randomized placement strategy optimized to shape the panel‐placement beam pattern.  By choosing optimized non-uniform, pseudo-random inter-panel spacings, we ensure that the virtual 1×10 array has a single dominant lobe in the desired direction, with the remaining sidelobes distributed as low‐amplitude “noise” (Fig.~\ref{fig:design_coherent_combining}).  When this optimized pattern multiplies the 1×8 ULA pattern, the resulting main beam remains sharp and stable in the desired direction, while the side lobes remain suppressed at all other angles.

Moreover, our evaluations show that using fewer panels with more elements per panel further improves robustness.  For instance, replacing sixteen 1×8 arrays with four 1×32 arrays achieves equivalent gain but yields a smoother beam profile: even if a panel’s peak direction deviates slightly, the aggregate beam still matches the ideal 1×32 ULA gain over the pointing error range. Thus, by decomposing the problem into per-panel beamforming and optimized inter-panel placement, \sysname enables grating-lobe‐free, coherent beamforming over kilometer-scale apertures.


\begin{figure}[t]
     \centering
     \begin{subfigure}[t]{0.24\textwidth}
         \centering
         \includegraphics[width=\textwidth]{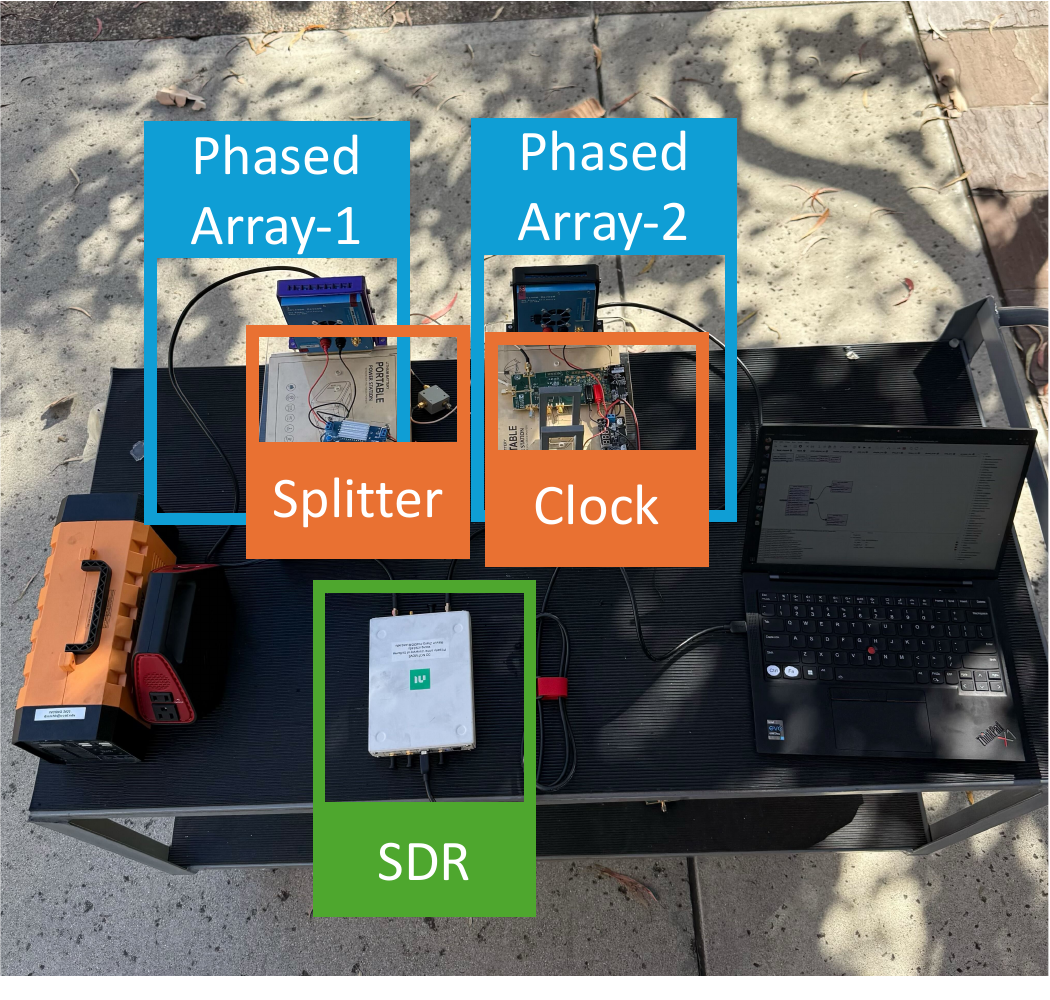}
         \caption{Transmitter/Receiver setup: two phased arrays sharing same clock and SDR for data collection}
         \label{fig:hardware_setup}
     \end{subfigure}
     \hfill
     \begin{subfigure}[t]{0.24\textwidth}
         \centering
         \includegraphics[width=\textwidth]{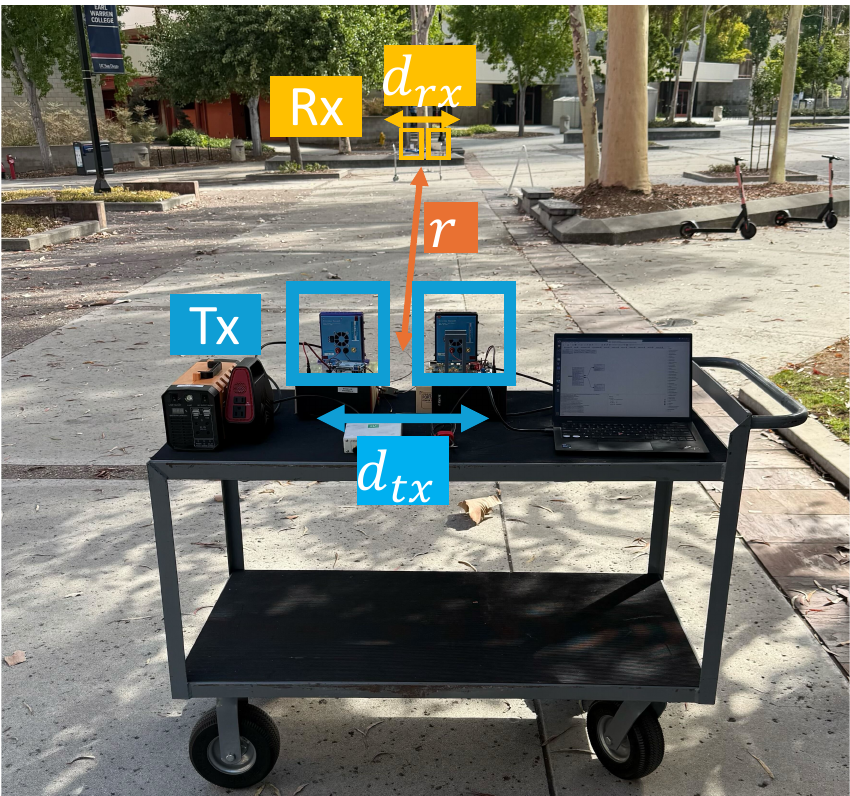}
         \caption{Experimental setup: Transmit and receive both with two phased in LoS. }
         \label{fig:hardware_experiments}
     \end{subfigure}
     \caption{\textbf{Hardware setup:} Illustrating experimental setup for hardware experiment. Varied phased arrays separations ($d_\mathrm{tx}$, ($d_\mathrm{rx}$)) and distance ($r$) between transmitter and receiver.}
    \label{fig:hardware_overall_setup}
    \vspace{-0.02\textheight}
\end{figure}

\begin{figure*}[!t]
     \centering
     \begin{subfigure}[t]{0.23\textwidth}
         \centering
         \includegraphics[width=\textwidth]{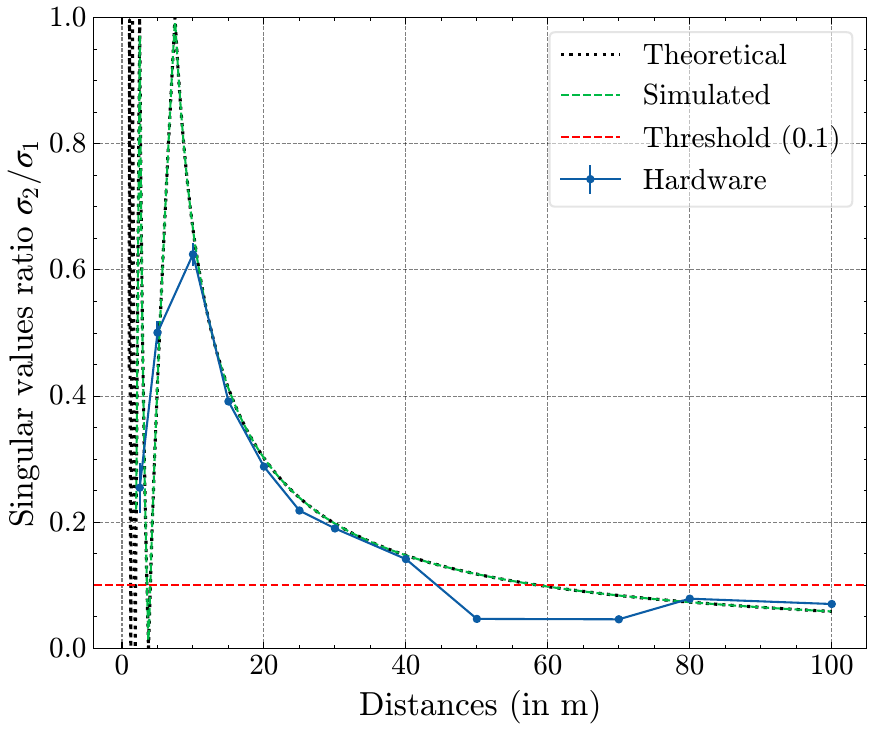}
         \caption{\centering $d_\mathrm{rx} = 20cm, d_\mathrm{tx} = 20cm$}
         \label{fig:hardware_results_dtx0.2_drx0.2}
     \end{subfigure}
     \hfill   
     \begin{subfigure}[t]{0.23\textwidth}
         \centering
         \includegraphics[width=\textwidth]{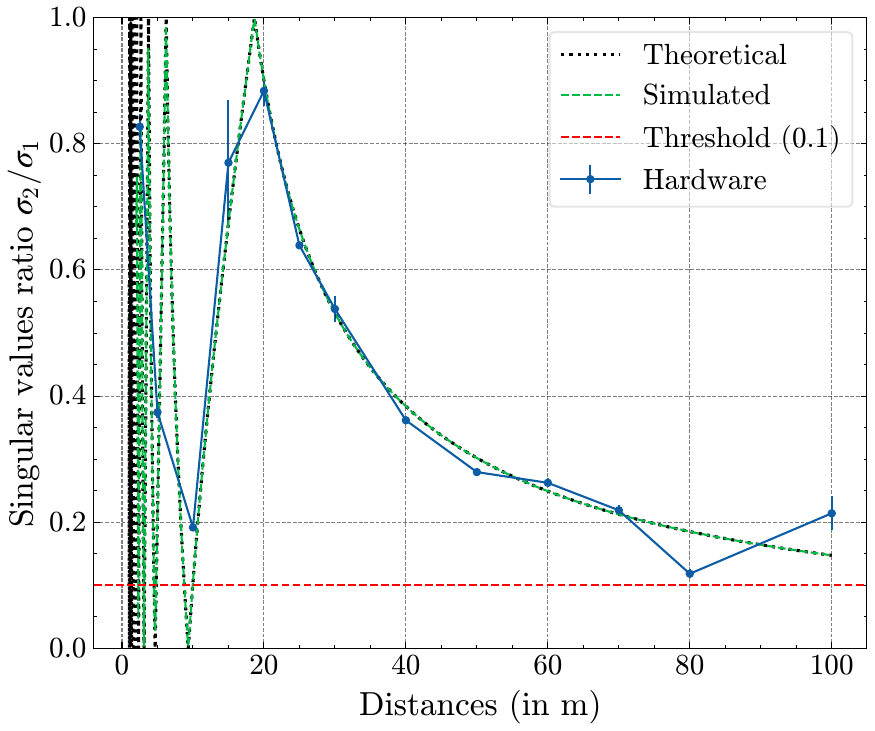}
         \caption{\centering $d_\mathrm{rx} = 20cm, d_\mathrm{tx} = 50cm$}
         \label{fig:hardware_results_dtx0.2_drx0.5}
     \end{subfigure}
     \hfill   
     \begin{subfigure}[t]{0.23\textwidth}
         \centering
         \includegraphics[width=\textwidth]{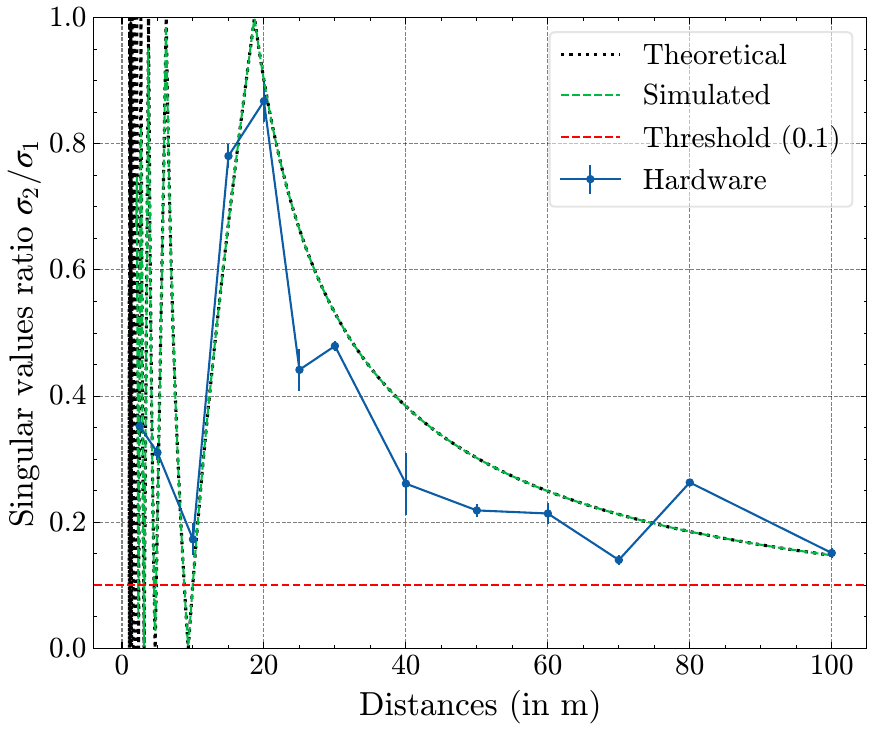}
         \caption{\centering  $d_\mathrm{rx} = 50cm, d_\mathrm{tx} = 20cm$}
         \label{fig:hardware_results_dtx0.5_drx0.2}
     \end{subfigure}
     \hfill   
     \begin{subfigure}[t]{0.23\textwidth}
         \centering
         \includegraphics[width=\textwidth]{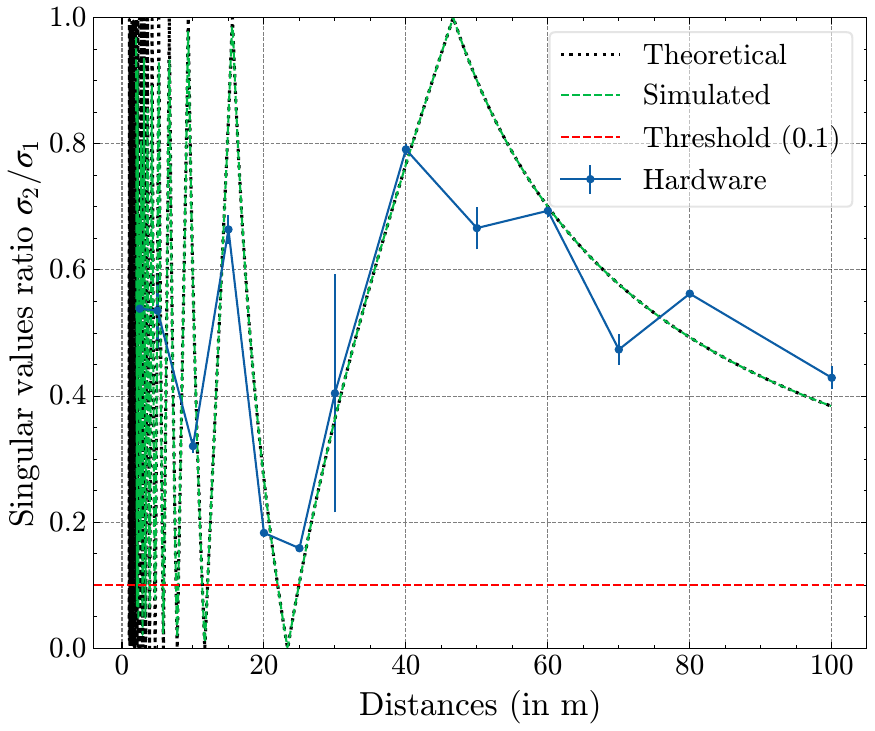}
         \caption{\centering  $d_\mathrm{rx} = 50cm, d_\mathrm{tx} = 50cm$}
         \label{fig:hardware_results_dtx0.5_drx0.5}
     \end{subfigure}
     \setlength{\belowcaptionskip}{-4pt}
     \caption{\textbf{Hardware results:} Demonstrating that our hardware results (blue), closely match with both theory (eq-4,5) and simulate channel (eq-3) for different transmit ($d_\mathrm{tx}$) and receive ($d_\mathrm{rx}$) antenna separations (2x2 Scenario).}
    \label{fig:hardware_exp_results}
    \vspace{-0.01\textwidth}
\end{figure*}

\section{Evaluations}\label{sec:evaluations}

To validate near-field MIMO feasibility in line-of-sight scenarios, we conducted both hardware experiments and simulations using a custom-built Python simulator.

\subsection{Hardware setup}
\label{subsec:eval_hardware}


\textbf{System Overview:}
The testbed consists of one transmitter (TX) node and one receiver (RX) node. Each node integrates two phase-coherent, vertically polarized 4 × 8 phased-array panels (64 elements apiece) that feed a single Ettus USRP B210 software-defined radio with a local oscillator at 3 GHz (Fig.\ref{fig:hardware_setup}). An external Analog Devices ADF5355 synthesizer provides a shared 6000 MHz reference clock distributed to the two panels through an RF splitter. Both arrays on a node connect to the B210’s two RF ports, ensuring port-to-port timing alignment inside the SDR while the common synthesizer guarantees frequency coherence across both arrays. This allows the phased arrays to operate at 27 GHz. All equipment is powered from a 12 V laboratory supply, and the B210 streams baseband I/Q data to a host PC over USB 3.0, where GNU Radio handles real-time signal generation on the TX side.

\begin{figure*}
     \centering
     \begin{subfigure}[t]{0.23\textwidth}
         \centering
         \includegraphics[width=\textwidth]{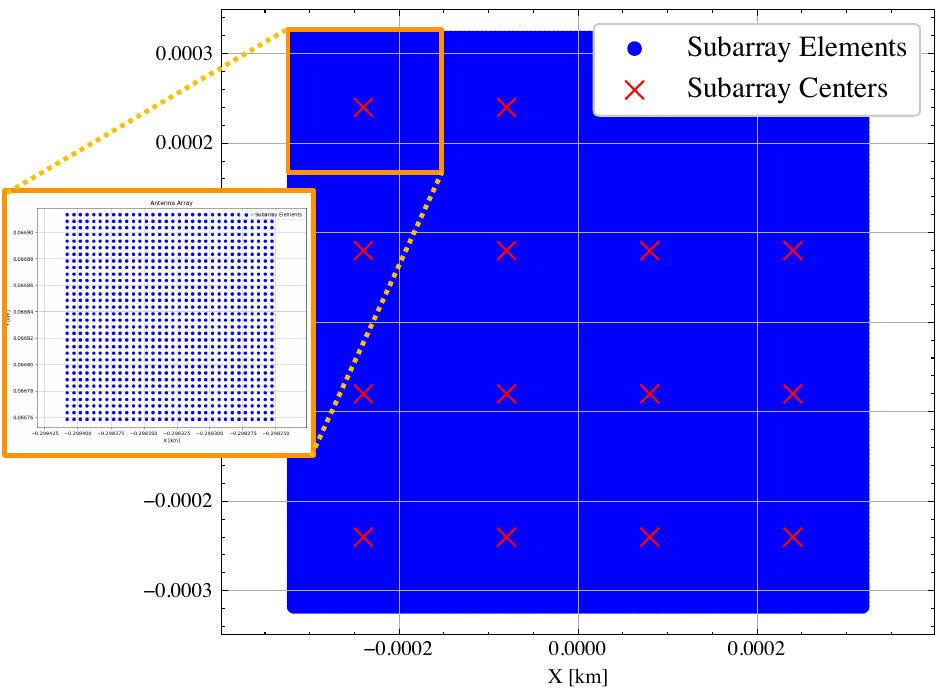}
         \caption{\centering 2D ULA with 128x128 antenna elements ($\lambda/2$ spacing).}
         \label{fig:eval_antenna_placement_ULA}
     \end{subfigure}
     \hfill   
     \begin{subfigure}[t]{0.24\textwidth}
         \centering
         \includegraphics[width=\textwidth]{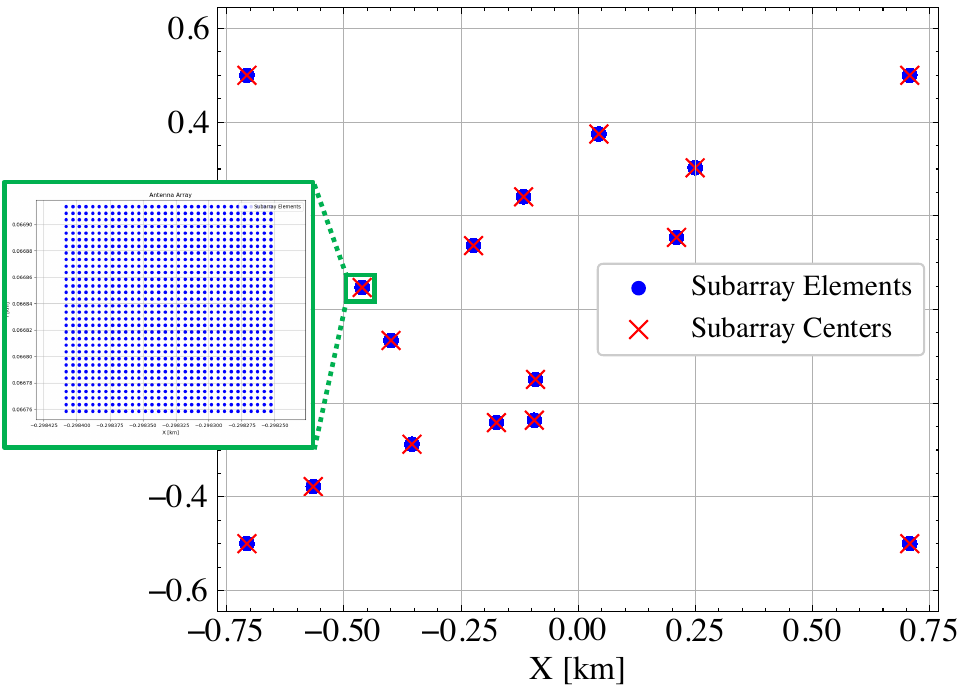}
         \caption{\centering \sysname: 16 phased arrays (32x32 UPAs) positions.}
         \label{fig:eval_antenna_placement_DSA}
     \end{subfigure}
     \hfill   
     \begin{subfigure}[t]{0.22\textwidth}
         \centering
         \includegraphics[width=\textwidth]{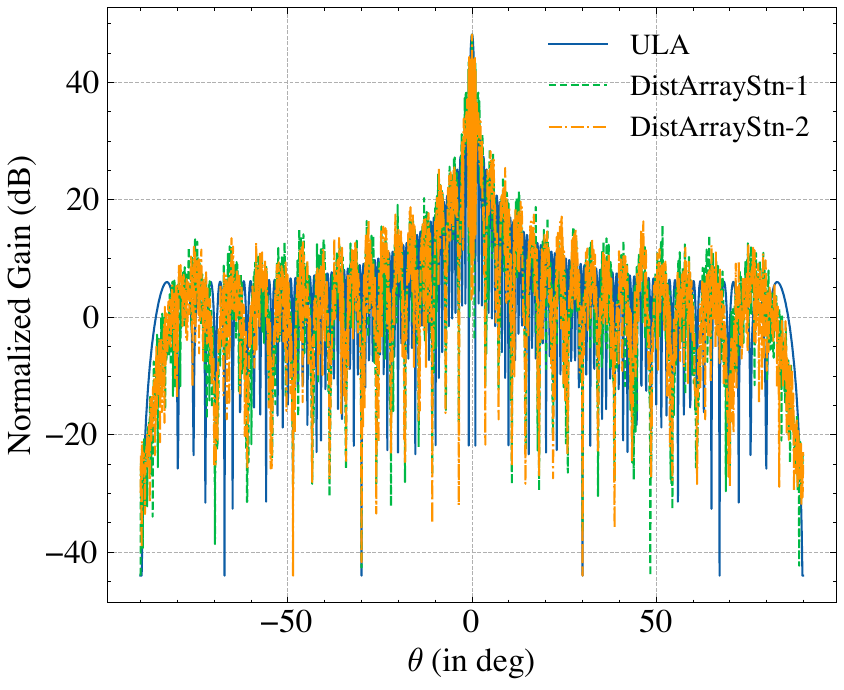}
         \caption{\centering Beam pattern vs $\theta$ (with 16 - 32x32 phased arrays).}
         \label{fig:beam_pattern_vs_theta_gain}
     \end{subfigure}
     \hfill   
     \begin{subfigure}[t]{0.22\textwidth}
         \centering
         \includegraphics[width=\textwidth]{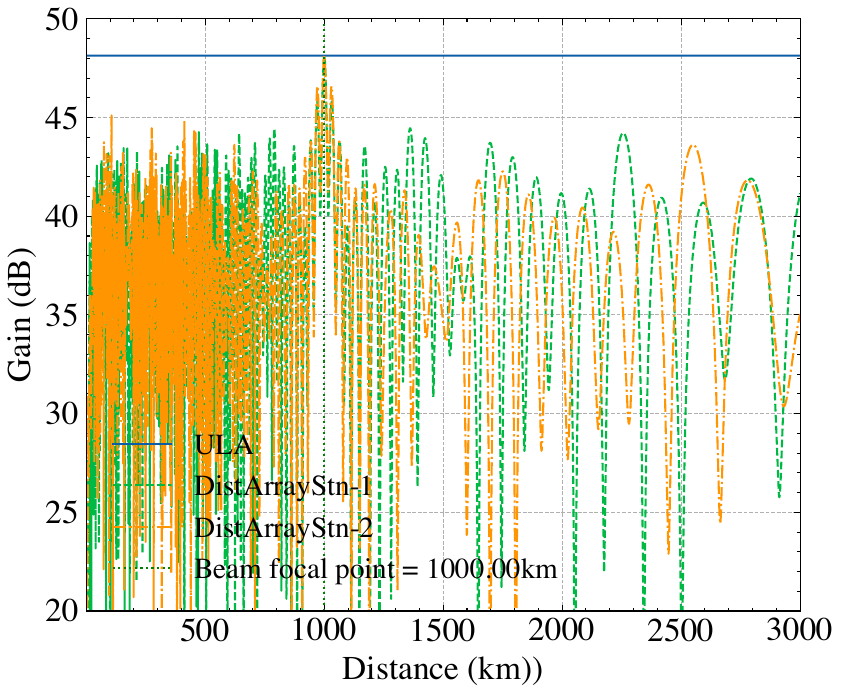}
         \caption{\centering Beam pattern vs distance (beam focusing in distance).}
         \label{fig:beam_pattern_vs_distance_gain}
     \end{subfigure}
     \setlength{\belowcaptionskip}{-4pt}
     \caption{\textbf{Simulation setup and results:} Antenna array placement for (a) 2D ULA with 128 x128 antenna elements and (b) \sysname with 16 32x32 distributed phased arrays in 1km x 1km grid.}
    \label{fig:antenna_placement}
    \vspace{-0.01\textwidth}
\end{figure*}

\textbf{Synchronization Strategy:}
With the B210’s internal VCXO bypassed, residual carrier-frequency offset (CFO) can exceed the tracking range of a standard 64-sample long-training sequence (LTS). To coarsely align the radios, the transmitter first emits a continuous 625 kHz tone; the receiver sweeps its center frequency until its FFT peak aligns with 625 kHz, reducing CFO to within a few kilohertz. Any remaining offset is removed digitally during packet processing, allowing the subsequent OFDM frames to be demodulated without cycle slips.

\textbf{Waveform Design:}
The link employs 2 × 2 MIMO OFDM. Each packet begins with a preamble tailored for both detection and channel estimation. Packet detection uses two copies of a 64-sample LTS, each preceded by a 32-sample cyclic prefix, for a total of 160 samples. Channel estimation then relies on two time-orthogonal 64-sample LTS blocks, again each with a 32-sample cyclic prefix. Overall, the preamble is 352 samples long (3 × 32 + 4 × 64). The data payload follows immediately, occupying the remaining OFDM symbols.

\textbf{Receiver Processing:}
A MATLAB supervisory script orchestrates acquisition on the RX host. It issues socket commands to a background Python worker that interfaces with UHD and captures 50 bursts per run. After each burst is recorded, MATLAB loads the raw samples, performs packet detection, corrects the fine CFO estimated from the preamble, and extracts the 2×2×64 channel-frequency-response matrix. The matrices from all 50 bursts are stored on disk for subsequent analysis of channel dynamics and beamforming performance.






\subsection{Hardware experiments}

We ran the experiment in an open outdoor area with a clear line of sight channel(Fig.\ref{fig:hardware_experiments}). Starting at 2.5 meters, we collected fifty packets and calculated the average singular value ratio at that range. We then increased the separation in five-meter steps up to thirty meters, after which we switched to ten-meter steps, upto 100m range.  
For each distance we tested four cases:
\begin{itemize}
    \item \textit{Case1:} Tx and Rx apertures: \(20\,\text{cm}\)
    \item \textit{Case2:} Tx aperture: \(50\,\text{cm}\); Rx aperture: \(20\,\text{cm}\)
    \item \textit{Case3:} Tx aperture: \(20\,\text{cm}\); Rx aperture: \(50\,\text{cm}\)
    \item \textit{Case4:} Tx and Rx apertures: \(50\,\text{cm}\)
\end{itemize}

In Fig.\ref{fig:hardware_exp_results}, the x-axis is the distance from the Transmitter to the receiver and the y-axis is the ratio of the second singular value to the first ($\frac{\sigma_1}{\sigma_0}$). To compute the singular value ratio for our hardware results, we first select one of the 64 frequency bins from the array, collapsing the data from a 64 x 2 × 2 × 50 array to a 2 × 2 × 50 array. For each of the 50 packets, we extracted 2 × 2 channel matrix and compute singular values using singular value decomposition (SVD). In Fig.\ref{fig:hardware_exp_results}, the circle indicate the mean and spread of each data point (most clearly visible on 7.d at 30 meters) indicates the standard deviation across all 50 packets. As shown, most data points have very small standard deviations, indicating a predictable channel.

For the theoretical plots, equations \ref{eq:sigma_eq} and \ref{eq:delta_standard} were used to determine the singular values (and their ratios) as a function of r for a given tx/rx aperture size and wavelength.

For the simulation, we applied \ref{eq:h_eq} to the exact antenna coordinates in a 2 × 2 MIMO arrangement to build the corresponding 2 × 2 channel matrix, and we then performed singular value decomposition on that matrix to extract the singular values and plot their ratios.

As shown in Fig.~\ref{fig:hardware_exp_results}, the hardware measurements closely match both theoretical predictions and simulation outputs, with minimal variance across packets. This consistency validates the feasibility of line-of-sight MIMO and the support of multiple simultaneous spatial streams to a single transceiver in the mmWave band.

\subsection{Simulation Framework for Satellite–Ground Links}
\label{subsec:eval_simulation_framework}
We developed a general-purpose Python simulator to analyze beamforming gains for arbitrary antenna configurations, including distributed phased arrays and conventional uniform arrays. The simulator accepts configurable antenna positions, satellite trajectories, and frequency parameters. For \sysname, we apply classic delay-and-sum weight computation to each panel pair, following the optimum beamforming approach in \cite{van2002optimum}.

\textit{Array Configurations:} We evaluated two deployment scenarios at 28 GHz (\(\lambda \approx 10.7\) mm):
\begin{itemize}
  \item \textbf{Uniform Planar Array (UPA)} - All \(128\times128\) elements grouped in a grid (Fig.~\ref{fig:eval_antenna_placement_ULA}).
  \item \textbf{Distributed Phased Arrays (DPA)} - 16 \(32\times32\) panels placed at the corners and randomly within a \(1.414\,\text{km}\times1\,\text{km}\) aperture (Fig.~\ref{fig:eval_antenna_placement_DSA}).
\end{itemize}

\begin{figure}[t]
     \centering
     \begin{subfigure}[t]{0.24\textwidth}
         \centering
         \includegraphics[width=\textwidth]{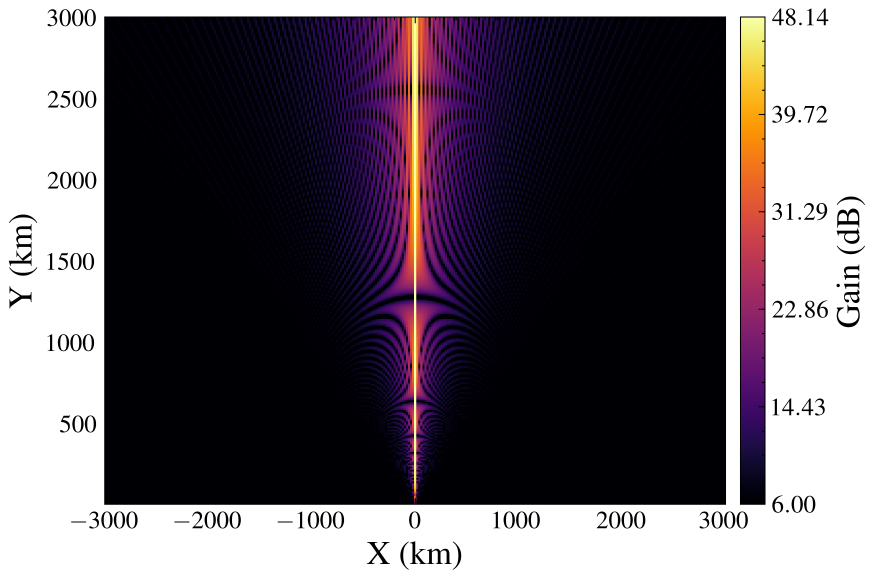}
         \caption{2D-ULA beamforming}
         \label{fig:2D_beamforming_ULA}
     \end{subfigure}
     \hfill
     \begin{subfigure}[t]{0.24\textwidth}
         \centering
         \includegraphics[width=\textwidth]{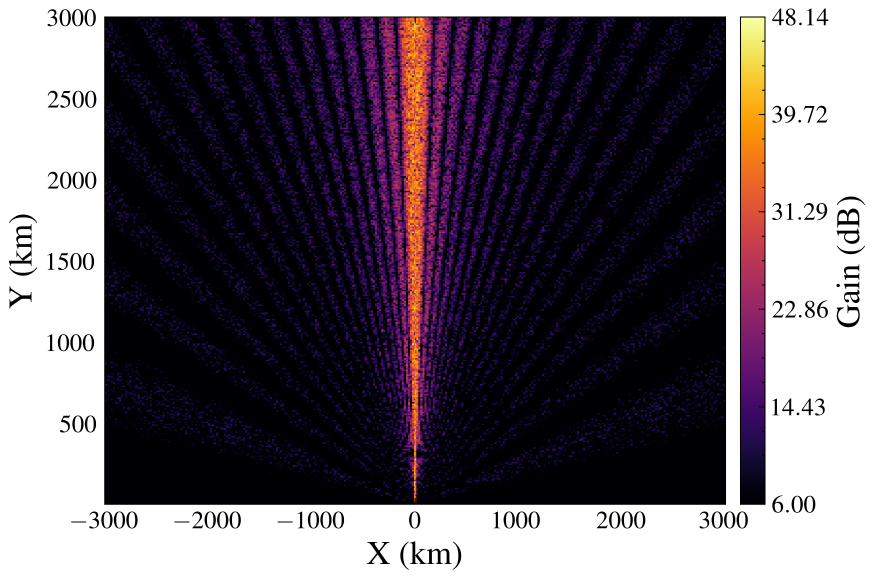}
         \caption{\sysname beamforming}
         \label{fig:2D_beamforming_DSA}
     \end{subfigure}
     \caption{
     An illustration 2D dimensional beamforming varying theta and distance. (a) 2D-ULA scenario (b) \sysname with phased arrays distributed in $1.414 km \times 1km$ area.   
     }
    \label{fig:2D_beam_patterns}
    \vspace{-0.02\textheight}
\end{figure}

\subsubsection{Coherent Combining}
\label{subsubsec:coherent_combining}

We evaluate \sysname’s ability to coherently combine sixteen \(32\times32\) phased-array panels against a benchmark \(128\times128\) uniform planar array (UPA), matching total element count. Beamforming weights for both systems are computed using classical delay-and-sum phase compensation \cite{van2002optimum}.

Figure~\ref{fig:beam_pattern_vs_theta_gain} compares array gain versus steering angle. The distributed phased arrays (green/orange curves for two different placement positions) achieve within 1–2 dB of the UPA (blue curve) across all angles, demonstrating near-parity in the angular beamforming performance. Figure~\ref{fig:beam_pattern_vs_distance_gain} plots array gain versus range. The UPA maintains a constant high gain at all distances, whereas \sysname localizes energy both angularly and radially—suppressing off-target lobes outside the intended range. In the reactive near-field (short ranges), gain patterns exhibit random fluctuations; beyond the transition to the radiative region, the patterns stabilize and closely track steering angles.  

Figure~\ref{fig:2D_beam_patterns} presents two-dimensional beam patterns over angle and range. The UPA exhibits a high-gain ridge at the steering angle that persists uniformly across range, whereas \sysname produces a localized gain peak that decays away from the focal range. This range localization not only preserves high on-axis gain for satellite links but also reduces off-axis interference.
In general, \sysname matches the high-gain beamforming of a monolithic UPA while additionally enabling distance-selective focusing, critical for interference mitigation in long-range satellite feeder links.




\begin{figure}[!t]
     \centering
     \begin{subfigure}[t]{0.23\textwidth}
         \centering
         \includegraphics[width=\textwidth]{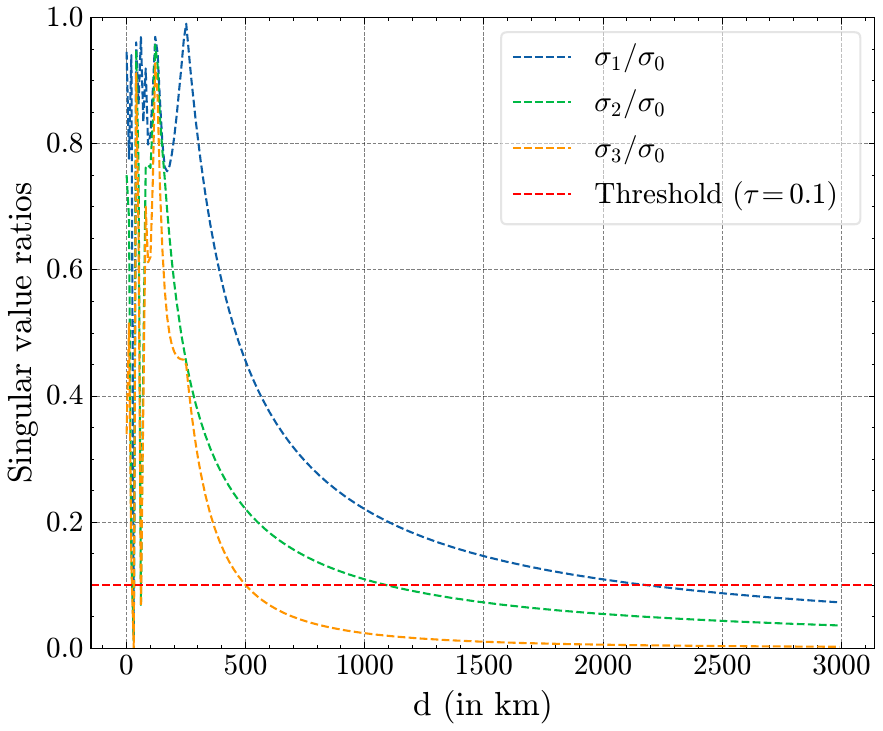}
         \caption{Singular value ratios: determines whether the channel is well-conditioned}
         \label{fig:dofs_eval_singular_values}
     \end{subfigure}
     \hfill
     \begin{subfigure}[t]{0.23\textwidth}
         \centering
         \includegraphics[width=\textwidth]{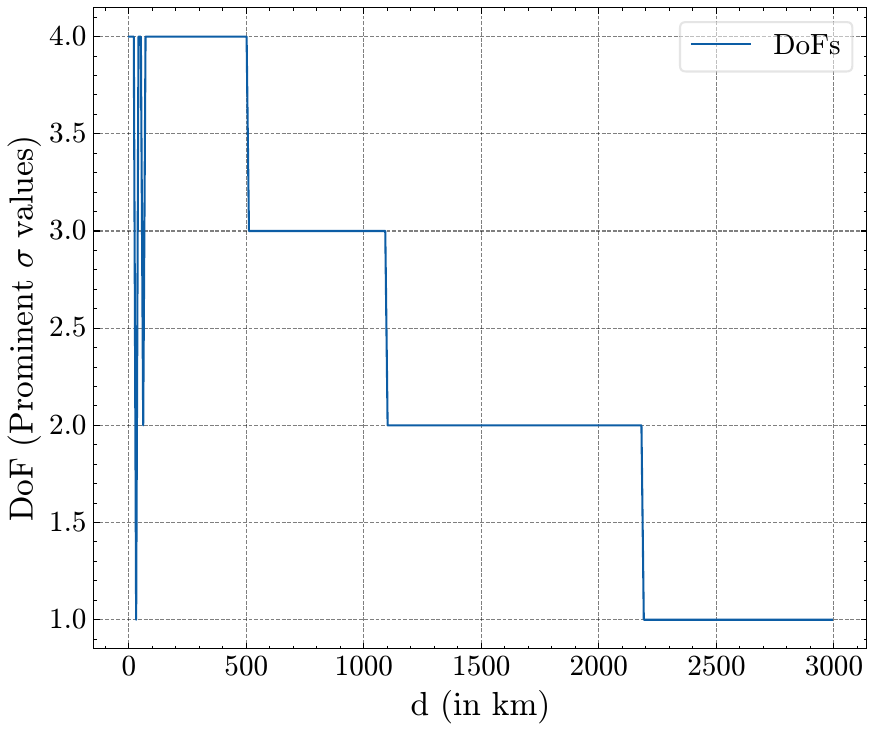}
         \caption{Channel rank: \#no of feasible simultaneous streams}
         \label{fig:dofs_eval_prominent_singular_values}
     \end{subfigure}
     \caption{Simulation results demonstrating \sysname’s LoS-MIMO capability with 16 \(32\times32\) phased-array panels distributed over a 1.414km×1km aperture and a four-element satellite array positioned at the corners of a 1.414m×1m grid.}
    \label{fig:dofs_simulation}
    \vspace{-0.01\textheight}
\end{figure}

\subsection{Line-of-Sight (LoS) MIMO Capability}
\label{subsec:eval_los_mimo_sim}

We further assess \sysname’s potential for LoS-MIMO. In the DPA scenario (Fig.~\ref{fig:eval_antenna_placement_DSA}), we position a four-element satellite array at the aperture corners of a \(1.414\,\text{km}\times1\,\text{km}\) grid. Using the channel model in Eq.~\ref{eq:h_eq}, we compute the channel singular values and degrees of freedom over ranges up to 3,000 km.  

Figure~\ref{fig:dofs_eval_singular_values} shows that the singular-value ratio \(\sigma_{\min}/\sigma_{\max}\) remains above threshold \(\tau=0.1\) up to 2,000 km, indicating stable channel invertibility. Figure~\ref{fig:dofs_eval_prominent_singular_values} plots the number of spatial streams: four streams are supported up to 500 km, three up to \(\sim\!1{,}000\) km, and two up to \(\sim\!2{,}000\) km.

These results confirm that \sysname not only matches high-gain beamforming but also unlocks multiple concurrent streams in clear-line-sight satellite feeder links.


\section{Conclusion}\label{sec:conclusion}

We have introduced \sysname, a distributed phased array ground station that coherently combines multiple small commercially available panels to achieve dish-class beamforming gain without the size and cost of a monolithic aperture. By spacing panels across a kilometer-scale aperture, \sysname operates in the radiative near-field to enable high-gain beam-focused links and unlock line-of-sight MIMO multiplexing, enabling multiple concurrent spatial streams on a feeder link. 

Our comprehensive evaluation, comprising analytical derivations, high‐fidelity simulations, and outdoor hardware experiments confirms that \sysname (1) matches or exceeds the gain of a 1.47 m parabolic plate through sixteen \(32\times32\) panels, (2) supports up to four LoS-MIMO streams at hundreds of kilometers and two streams beyond 2,000 km, and (3) exhibits strong agreement between theory, simulation, and measurement with minimal variance. These results demonstrate a practical and cost-effective way to scale the LEO ground station capacity. 

Future work will explore adaptive panel placement algorithms, real‐time over‐the‐air calibration across heterogeneous rooftops, and integration with multi‐satellite support to further enhance throughput, flexibility, and resilience in next‐generation satellite networks.

\section{Acknowledgements}
We thank the WCSNG team members at UC San Diego for their valuable feedback. 
This research was partially supported by the National Science Foundation under grants 2232481 and 2211805.

\appendix
\section{Derivation of Singular Values for Unit‐Modulus $2\times2$ Matrix}
Let
\[
H=\begin{pmatrix}
 h_0 & h_1\\
 h_2 & h_3
\end{pmatrix},
\quad
h_i = e^{-j\theta_i},\quad|h_i|=1.
\]
The Frobenius norm is
\[
\|H\|_F^2 = \sum_{i=0}^3 |h_i|^2 = 4.
\]
The determinant is
\[
\det H = h_0 h_3 - h_1 h_2 = e^{-j(\theta_0+\theta_3)} - e^{-j(\theta_1+\theta_2)}.
\]
Define
\[
\alpha = \theta_0 + \theta_3,\quad
\beta = \theta_1 + \theta_2,\quad
\Delta = \alpha - \beta.
\]
Then
\[
|\det H|
= \bigl|e^{-j\alpha}-e^{-j\beta}\bigr|
= 2\Bigl|\sin\frac{\Delta}{2}\Bigr|.
\]
The squared singular values satisfy the characteristic equation
\[
\sigma^4 - \|H\|_F^2\,\sigma^2 + |\det H|^2 = 0,
\]
so
\[
\sigma_{1,2}^2
= \frac{\|H\|_F^2}{2}
  \pm \frac{1}{2}\sqrt{\|H\|_F^4 - 4|\det H|^2}
= 2 \pm 2\Bigl|\cos\frac{\Delta}{2}\Bigr|.
\]
Hence the singular values are
\[
\sigma_{1,2} = \sqrt{2 \pm 2\Bigl|\cos\tfrac{\Delta}{2}\Bigr|}.
\]

\bibliographystyle{IEEEtran}
\bibliography{main}

\begin{thebibliography}{10}
\providecommand{\url}[1]{#1}
\csname url@samestyle\endcsname
\providecommand{\newblock}{\relax}
\providecommand{\bibinfo}[2]{#2}
\providecommand{\BIBentrySTDinterwordspacing}{\spaceskip=0pt\relax}
\providecommand{\BIBentryALTinterwordstretchfactor}{4}
\providecommand{\BIBentryALTinterwordspacing}{\spaceskip=\fontdimen2\font plus
\BIBentryALTinterwordstretchfactor\fontdimen3\font minus \fontdimen4\font\relax}
\providecommand{\BIBforeignlanguage}[2]{{%
\expandafter\ifx\csname l@#1\endcsname\relax
\typeout{** WARNING: IEEEtran.bst: No hyphenation pattern has been}%
\typeout{** loaded for the language `#1'. Using the pattern for}%
\typeout{** the default language instead.}%
\else
\language=\csname l@#1\endcsname
\fi
#2}}
\providecommand{\BIBdecl}{\relax}
\BIBdecl

\bibitem{SpaceX_wiki}
wiki, ``\textit{SpaceX.}'' \url{https://en.wikipedia.org/wiki/SpaceX}, Aug 2025.

\bibitem{Planet_Labs_wiki}
------, ``\textit{Planet Labs.}'' \url{https://en.wikipedia.org/wiki/Planet_Labs}, Aug 2025.

\bibitem{Project_Kuiper_Wiki}
------, ``\textit{Project Kuiper.}'' \url{https://en.wikipedia.org/wiki/Project_Kuiper}, Aug 2025.

\bibitem{Eutelsat_OneWeb_wiki}
------, ``\textit{Eutelsat OneWeb.}'' \url{https://en.m.wikipedia.org/wiki/Eutelsat_OneWeb#OneWeb_satellite_constellation}, Aug 2025.

\bibitem{fcc_spacex_LA_gndstn}
FCC, ``\textit{RADIO STATION AUTHORIZATION},'' \url{https://fcc.report/IBFS/SES-LIC-20200428-00458/3872480.pdf}, Feb 2021.

\bibitem{fcc_spacex_gnd_stn}
fcc, ``\textit{Application for Fixed Satellite Service by SpaceX Services, Inc.}'' \url{https://fcc.report/IBFS/SES-AMD-20230525-01127}, Aug 2025.

\bibitem{he2021review}
G.~He, X.~Gao, L.~Sun, and R.~Zhang, ``A review of multibeam phased array antennas as leo satellite constellation ground station,'' \emph{IEEE Access}, vol.~9, pp. 147\,142--147\,154, 2021.

\bibitem{balanis2016antenna}
C.~A. Balanis, \emph{Antenna theory: analysis and design}.\hskip 1em plus 0.5em minus 0.4em\relax John wiley \& sons, 2016.

\bibitem{ortbital_rot_per_sec}
C.~O. Systems, ``\textit{2.4AEBP-3m Elevation-Over-Azimuth Antenna Positioner},'' \url{https://www.orbitalsystems.com/wp-content/uploads/2020/06/2.4AEBP-3m-Data-Sheet-D.01.pdf}, June 2025.

\bibitem{weerackody2006motion}
V.~Weerackody and L.~Gonzalez, ``Motion induced antenna pointing errors in satellite communications on-the-move systems,'' in \emph{2006 40th Annual Conference on Information Sciences and Systems}.\hskip 1em plus 0.5em minus 0.4em\relax IEEE, 2006, pp. 961--966.

\bibitem{geng2021correction}
M.~Geng-Jun, X.~Bin-Bin, W.~Na, and W.~Zhao-Jun, ``Correction method of antenna pointing error caused by the main reflector deformation,'' \emph{Chinese Astronomy and Astrophysics}, vol.~45, no.~2, pp. 236--251, 2021.

\bibitem{wiki_phased_array}
Wikipedia, ``\textit{Phased array},'' \url{https://en.wikipedia.org/wiki/Phased_array}, July 2025.

\bibitem{kang2025nmap}
J.-M. Kang, ``Nmap-net: Deep learning-aided near-field multi-beamforming design and antenna position optimization for xl-mimo communications,'' \emph{IEEE Internet of Things Journal}, 2025.

\bibitem{huang2025symmetric}
K.~Huang, J.~Guan, Y.~Wang, and X.~Luo, ``A symmetric beam training method for near-field xl-mimo systems,'' in \emph{2025 6th International Conference on Electrical, Electronic Information and Communication Engineering (EEICE)}.\hskip 1em plus 0.5em minus 0.4em\relax IEEE, 2025, pp. 1491--1496.

\bibitem{li2025blind}
J.~Li, Q.~Chen, and F.~Xi, ``A blind super-resolution method for near-field channel estimation with angle-range recovery,'' in \emph{ICASSP 2025-2025 IEEE International Conference on Acoustics, Speech and Signal Processing (ICASSP)}.\hskip 1em plus 0.5em minus 0.4em\relax IEEE, 2025, pp. 1--5.

\bibitem{liu2025gradient}
S.~Liu, H.~Wang, S.~Li, J.~Wang, H.~Li, and F.~Wang, ``Gradient descent based polarization channel estimation in extremely largescale mimo systems,'' in \emph{2025 IEEE 17th International Conference on Computer Research and Development (ICCRD)}.\hskip 1em plus 0.5em minus 0.4em\relax IEEE, 2025, pp. 155--160.

\bibitem{lei2025near}
H.~Lei, J.~Zhang, Z.~Wang, B.~Ai, and E.~Bjornson, ``Near-field user localization and channel estimation for xl-mimo systems: Fundamentals, recent advances, and outlooks,'' \emph{IEEE Wireless Communications}, 2025.

\bibitem{zhou2025super}
C.~Zhou, C.~You, S.~Shi, J.~Zhou, and C.~Wu, ``Super-resolution wideband beam training for near-field communications with ultra-low overhead,'' \emph{IEEE Internet of Things Journal}, 2025.

\bibitem{liu2025dynamic}
M.~Liu, M.~Li, R.~Liu, and Q.~Liu, ``Dynamic hybrid beamforming designs for elaa near-field communications,'' \emph{IEEE Journal on Selected Areas in Communications}, 2025.

\bibitem{zhou2025ris}
Q.~Zhou, J.~Zhao, K.~Cai, and Y.~Zhu, ``Ris-assisted beamfocusing in near-field iot communication systems: A transformer-based approach,'' \emph{IEEE Internet of Things Journal}, 2025.

\bibitem{sun2025near}
P.~Sun and B.~Wang, ``Near-field beam-focusing for integrated sensing and communication systems,'' \emph{IEEE Internet of Things Journal}, 2025.

\bibitem{cui2022near}
M.~Cui, Z.~Wu, Y.~Lu, X.~Wei, and L.~Dai, ``Near-field mimo communications for 6g: Fundamentals, challenges, potentials, and future directions,'' \emph{IEEE Communications Magazine}, vol.~61, no.~1, pp. 40--46, 2022.

\bibitem{parvini2025tutorial}
M.~Parvini, B.~Banerjee, M.~Q. Khan, T.~Mewes, A.~Nimr, and G.~Fettweis, ``A tutorial on wideband xl-mimo: Challenges, opportunities and future trends,'' \emph{IEEE Open Journal of the Communications Society}, 2025.

\bibitem{sarris2005maximum}
I.~Sarris and A.~R. Nix, ``Maximum mimo capacity in line-of-sight,'' in \emph{2005 5th International Conference on Information Communications \& Signal Processing}.\hskip 1em plus 0.5em minus 0.4em\relax IEEE, 2005, pp. 1236--1240.

\bibitem{jiang2022design}
K.~Jiang, X.~Wang, Y.~Jin, A.~Saleem, and G.~Zheng, ``Design and analysis of high-capacity mimo system in line-of-sight communication,'' \emph{Sensors}, vol.~22, no.~10, p. 3669, 2022.

\bibitem{fernandez2024constellation}
R.~Fernandez, Y.~Ma, and R.~Tafazolli, ``Constellation with optimal leo spacing for satellite-to-mobile downlink communication,'' in \emph{2024 IEEE-APS Topical Conference on Antennas and Propagation in Wireless Communications (APWC)}.\hskip 1em plus 0.5em minus 0.4em\relax IEEE, 2024, pp. 114--118.

\bibitem{pan2023pmsat}
H.~Pan, L.~Qiu, B.~Ouyang, S.~Zheng, Y.~Zhang, Y.-C. Chen, and G.~Xue, ``Pmsat: Optimizing passive metasurface for low earth orbit satellite communication,'' in \emph{Proceedings of the 29th Annual International Conference on Mobile Computing and Networking}, 2023, pp. 1--15.

\bibitem{chang2024smart}
M.-L. Chang, D.-B. Lin, H.-T. Rao, H.-Y. Lin, and H.-T. Chou, ``Smart transfer planer with multiple antenna arrays to enhance low earth orbit satellite communication ground links.'' \emph{Electronics (2079-9292)}, vol.~13, no.~17, 2024.

\bibitem{merino2022phased}
I.~Merino-Fernandez, S.~L. Khemchandani, J.~Del~Pino, and J.~Saiz-Perez, ``Phased array antenna analysis workflow applied to gateways for leo satellite communications,'' \emph{Sensors}, vol.~22, no.~23, p. 9406, 2022.

\bibitem{adomnitei2024phased}
C.-I. Adomnitei, C.-E. Lesanu, A.~Done, E.~Coca, and A.~Lavric, ``Phased antenna arrays, software defined radio and artificial intelligence: Advancing leo satellite communications,'' \emph{Advances in Electrical and Computer Engineering}, vol.~24, no.~4, pp. 57--64, 2024.

\bibitem{fcc_spacex_phased_arrays}
FCC, ``\textit{Spacex Purpose of Experiment},'' \url{https://apps.fcc.gov/els/GetAtt.html?id=259301}, July 2025.

\bibitem{kuiper_terminals}
Amazon, ``\textit{Here’s your first look at Project Kuiper’s low-cost customer terminals},'' \url{https://www.aboutamazon.com/news/innovation-at-amazon/heres-your-first-look-at-project-kuipers-low-cost-customer-terminals}, March 2023.

\bibitem{fcc_oneweb_phased_arrays}
FCC, ``\textit{ONEWEB NON-GEOSTATIONARY SATELLITE SYSTEM (LEO)},'' \url{https://fcc.report/IBFS/SAT-MPL-20200526-00062/2379706.pdf}, July 2025.

\bibitem{extereme_waves_pa}
E.~Waves, ``\textit{SATCOM Ground Terminals},'' \url{https://www.extreme-waves.com/satcom}, July 2025.

\bibitem{wiki_microstrip_antenna}
Wikipedia, ``\textit{Microstrip antenna},'' \url{https://en.wikipedia.org/wiki/Microstrip_antenna}, March 2025.

\bibitem{mailloux2017phased}
R.~J. Mailloux, \emph{Phased array antenna handbook}.\hskip 1em plus 0.5em minus 0.4em\relax Artech house, 2017.

\bibitem{hansen2009phased}
R.~C. Hansen, \emph{Phased array antennas}.\hskip 1em plus 0.5em minus 0.4em\relax John Wiley \& Sons, 2009.

\bibitem{liu2023near}
Y.~Liu, Z.~Wang, J.~Xu, C.~Ouyang, X.~Mu, and R.~Schober, ``Near-field communications: A tutorial review,'' \emph{IEEE Open Journal of the Communications Society}, vol.~4, pp. 1999--2049, 2023.

\bibitem{Rohde_Schwarz_MIMO}
R.~. Schwarz, ``\textit{Assessing a MIMO Channel - White Paper.}'' \url{https://scdn.rohde-schwarz.com/ur/pws/dl_downloads/dl_application/application_notes/8nt1/8NT1_0e_Assessing_MIMO_Ch.pdf}, Aug 2025.

\bibitem{van2002optimum}
H.~L. Van~Trees, \emph{Optimum array processing: Part IV of detection, estimation, and modulation theory}.\hskip 1em plus 0.5em minus 0.4em\relax John Wiley \& Sons, 2002.

\end{thebibliography}

\end{document}